\documentclass[twocolumn,english,prx]{revtex4-1}
\usepackage[latin9]{inputenc}
\usepackage{verbatim}
\usepackage{units}
\usepackage{amsmath}
\usepackage{amssymb}
\usepackage{graphicx}
\usepackage{esint}



\usepackage{babel}
\begin{document}
\global\long\def\ket#1{\left|#1\right\rangle }

\global\long\def\bra#1{\left\langle #1\right|}

\global\long\def\braket#1#2{\left\langle #1\left|#2\right.\right\rangle }

\global\long\def\ketbra#1#2{\left|#1\right\rangle \left\langle #2\right|}

\global\long\def\braOket#1#2#3{\left\langle #1\left|#2\right|#3\right\rangle }

\global\long\def\mc#1{\mathcal{#1}}

\global\long\def\nrm#1{\left\Vert #1\right\Vert }

\title{Additional energy-information relations in thermodynamics of small
systems}

\author{Raam Uzdin}

\affiliation{Technion - Israel Institute of Technology, Haifa 3200003, Israel,
Faculty of Chemistry.}

\affiliation{University of Maryland, College Park, Maryland 20742, USA, Department
of Chemistry and Biochemistry}
\begin{abstract}
The Clausius inequality (CI) form of the second law of thermodynamics
relates information changes (entropy) to changes in the first moment
of the energy (heat and indirectly also work). Are there similar relations
between other moments of the energy distribution, and other information
measures, or is the Clausius inequality a one of a kind instance of
the energy-information paradigm? If there are additional relations,
can they be used to make predictions on measurable quantities? Changes
in the energy distribution beyond the first moment (average heat or
work) are especially important in small systems which are often very
far from thermal equilibrium. The generalized Clausius inequalities
(GCI's), here derived, provide positive answers to the two questions
above and add another layer to the fundamental connection between
energy and information. To illustrate the utility of the new GCI's,
we find scenarios where the GCI's yield tighter constraints on performance
(e.g. in thermal machines) compared to the second law. To obtain the
GCI's we use the Bregman divergence - a mathematical tool found to
be highly suitable for energy-information studies. The quantum version
of the GCI's provides a thermodynamic meaning to various quantum coherence
measures. It is intriguing to fully map the regime of validity of
the GCI's and extend the present results to more general scenarios
including continuous systems and particles exchange with the baths.
\end{abstract}
\maketitle

\section{Introduction}

Thermodynamics is a remarkable theory. It was originally conceived
to describe practical limitations of steam engines, and now it is
one of the pillars of theoretical physics with applications in countless
systems and scenarios. As Einstein said ``It is the only physical
theory of universal content, which I am convinced, that within the
framework of applicability of its basic concepts will never be overthrown''.
It is now established that basic thermodynamic laws such as the Clausius
inequality (second law) hold even when the system is composed of a
single particle with only few energy levels and the evolution is non
classical \cite{alicki79,Sagawa2012second,PeresBook,Esposito2011EPL2Law}.
Consequently, the Carnot efficiency limit holds for arbitrary small
and/or quantum heat machines. Nevertheless, this does not exclude
the appearance of quantum effects in microscopic heat machines \cite{EquivPRX,gelbwaser2014heat,MitchisonHuber2015CoherenceAssitedCooling}.
Even without quantum interference or entanglement the thermodynamics
of small systems is fascinating. Small systems like biological machines
typically operate far from equilibrium and may be subjected to strong
thermal fluctuations. For example, thermodynamics has been applied
to the study of biological replication of DNA \cite{GaspardCopyDNA,JarzynskiOnGaspard}.
More generally, non-equilibrium statistical mechanics, and stochastic
thermodynamics have been the subject of intensively study in recent
years (see \cite{Seifert2012StochasticReview,harris2007fluctuationReview,Jarzynski2011equalitiesReview}
and references therein). A single ion heat engine \cite{rossnagelIonEngExp}
and a multiple ion refrigerator \cite{Sing3ionEng2017} have recently
been experimentally demonstrated, and there are various suggestions
for heat machine realization in superconducting circuits \cite{PekolaSCengine,campisi2014FT_SolidStateExp},
optomechanics \cite{Kurizki2015workOptoMech,ZhangOptoMechEng} and
cavity QED \cite{Mitchison2016cavityQED}.

Thermodynamics has been traditionally applied to macroscopic objects
where deviations from averaged quantities (even outside equilibrium)
are too small to be measured or to be of any practical interest. With
the growing experimental capabilities in the microscopic realm, there
is a growing motivation to consider fluctuations from a thermodynamic
point of view and go beyond the first moment of the energy distribution.
Are there second laws for other quantities? If there are such laws,
we ask if they can be expressed in energy-information form and maintain
the structure of the standard second law. 

The relation between energy and information has led to a deep understanding
of the foundations of thermodynamics. Just to mention a few examples:
Maxwell demon, Szilard engine and Landauer erasure principle. The
Clausius inequality (CI) presents the energy information relationship
in a clear and concise way 
\begin{equation}
\Delta S-\int\delta Q/T(t)\ge0,\label{eq: CI alpha=00003D1}
\end{equation}
$Q$ is the heat exchanged with a bath at temperature $T(t)$, and
$S$ is the entropy of the system. The heat and entropy relation is
used daily in the study of thermal interactions. For example, in a
first order phase transition the latent heat is associated with the
disorder difference of the two phases. The Clausius inequality (\ref{eq: CI alpha=00003D1})
is one of the most versatile forms of the second law. It applies to
non-periodic processes, to multiple heat baths (as needed for heat
machines), and also for states that are initially and/or finally far
from thermal equilibrium \cite{JarzynskiMicroscopicClausius}. Moreover,
as mentioned above, the CI holds even in the quantum microscopic realm. 

In contrast to a macroscopic fluid at equilibrium, at the microscopic
scale the system typically does not have a classical equation of state
with just a few thermodynamic variables. The entropy that appears
in the CI in such a case is the von Neumann entropy \cite{PeresBook}
of the system, and it is defined regardless of equilibrium or an equation
of state. Up to Sec. \ref{sec: Quantum gCI}, we deal only with statistical
mixtures of energy eigenstates, so the von Neumann entropy reduces
to the Shannon entropy in the energy basis of the system $S=\sum_{j}-p_{j}\ln p_{j}$,
as in the framework of stochastic thermodynamics \cite{Seifert2012StochasticReview}.

In recent years, the second law has been explored and extended in
two very different frameworks. The first is stochastic thermodynamics
(and non-equilibrium statistical mechanics). The Jarzynski fluctuation
relations for work \cite{quan2008quantumFluctTheorem,Jarzynski2011equalitiesReview},
can be viewed as a generalization of the second law (in certain scenarios)
which is applicable to higher energy moments. This approach has been
successfully applied to heat machines as well \cite{campisi14,campisi2014FT_SolidStateExp}.
The results in \cite{campisi14,campisi2014FT_SolidStateExp} are important
and interesting, but they do not relate changes in information to
changes in energy. The other framework that can be viewed as an extension
of the second law is thermodynamic resource theory \cite{Goold2015review,horodecki2013fundamental,GourRTreview,LostaglioRudolphCohConstraint}.
In this framework properties of completely positive maps are used
to construct monotones that must decrease under thermal interactions
with a single bath. Despite the appealing elegance of this framework
it has a major drawback: these monotones are so far not related to
observable quantities. Nonetheless, there are interesting insights
arising from this framework (e.g. \cite{MatteoIndependenceResource}).
More important for the present work is the fact that resource theory
is not formulated in terms of information and energy. A more extensive
discussion on resource theory appears in Appendix IV. 

In this paper we present a third way of extending the second law.
Our approach is based on the energy-information paradigm and has the
same logic and underlying structure as the standard second law. Our
results clearly show that such additional energy-information relations
exist. The results presented here should be further extended and explored.
Yet, we emphasize that even this first study provides new predictions
and significantly extends our understanding of the interplay between
energy, information, and the mathematical framework that connects
them. A summary of the three approaches for extending the second law
is given in Table 1.
\begin{table}
\includegraphics[width=8.6cm]{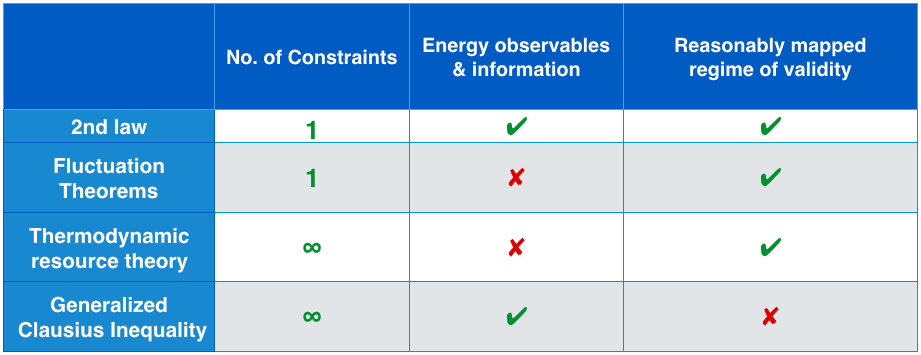}\caption{Comparison between different frameworks that predict constraints on
thermodynamic processes. The standard second law is not just a mathematical
constraint, it has an energy-entropy structure. Among the known extensions
of the second law only the generalized Clausius inequalities here
derived provides new energy-information relations. Other features
are compared in the text.}

\end{table}

Finally, we wish to give one more motivating arguments for the study
of additional second laws. Consider the following scenario: a system
with a time-independent Hamiltonian $H$ is connected to a thermal
bath and reaches a thermal state with an average energy $\left\langle H\right\rangle _{f}$.
The heat in this case is determined by the change in the average energy
with respect to the initial state $Q=\left\langle H\right\rangle _{f}-\left\langle H\right\rangle _{0}$.
When the initial state is not thermal but satisfies $\left\langle H\right\rangle _{0}=\left\langle H\right\rangle _{f}$
something puzzling takes place from the thermodynamic point of view.
It is clear that the bath has changed the energy probability distribution
of the system (despite the fact that the average has not changed).
We ask: 1) Since the system has changed, the bath must have changed
as well, is it possible to thermodynamically quantify the change in
the bath when $Q=0$? 2) A change in the energy probability distribution
of the system, implies that some of the energy moments have changed
as well. Is it possible to formulate a thermodynamic framework for
the change in energy moments other than the first? Even though there
seems to be no immediate reason to assume there are thermodynamic
answers to these questions, this paper provides a possible answer
by formulating generalized Clausius inequalities for higher order
energy moments.

Changes in the higher moments of the energy are not only important
for understanding the system dynamics far from equilibrium. They are
also important for understanding the back action on heat reservoirs
with finite heat capacity. By studying the changes in higher moments
of the bath, it is possible to quantify to what extent the bath has
deviated from equilibrium by interacting with the system (when the
heat capacity of the bath is not infinite as in the ideal case). For
example, for a thermal state in the bath, we expect a certain relation
between the first and second moment of the energy. This relation still
holds if the bath is heated to a different thermal state. However,
if energy entered the bath, but the bath does not relax to equilibrium
(e.g. because it is too small), then the thermal relation between
the moments will no longer hold. We return to this point later on
when discussing the impact on the bath.

The term 'generalized Clausius inequality' has already appeared in
\cite{DeffnerLutzRelEntBures}. There, the CI is written using the
relative entropy, and a divergence inequality is used to obtain an
'improved CI'. The zero in one of the sides of the CI is replaced
by some positive number related to the Bures length. However, the
physical quantities and the information measures are still the same
as in the standard inequality. In contrast, in this study we generalized
the energy-information measures to other quantities.

\section{Main findings\label{sec: Main-findings}}

In this section we present the generalized CI's in their simplest
form and not in their more general form derived in Appendix I. In
the main text we deal with cases where only one bath is connected
to the system at a given time. However several baths can be connected
sequentially in time so that the temperature of the bath is time dependent.
That is instead of the notation $\sum\beta_{k}Q_{k}$ in the CI we
shall use the notation $\intop\frac{\delta Q}{T(t)}$ for sequential
connection to multiple baths at different temperatures. The generalization
to simultaneous connection to several baths is discussed in Appendix
I. 

Let the system be composed of a finite set of states $j=1,..,N$ whose
energies are $\{E_{j}\}_{j=1}^{N}$. The probability to be in a state
'$j$' is $p_{j}$ (to separate the stochastic part of the paper from
the quantum part, we use quantum notations only in Sec. \ref{sec: Quantum gCI}).
The energy probability distribution of a thermal state with inverse
temperature $\beta=1/T$ is $p_{\beta,j}=e^{-\beta(E_{j}-F)}$, $F=-\frac{1}{\beta}\ln\sum_{j}e^{-\beta E_{j}}=-\frac{1}{\beta}\ln Z$
is the standard free energy, and $Z$ is the partition function. As
in the stochastic thermodynamics framework \cite{Seifert2012StochasticReview}
the energy levels can be varied in time, and the system (or parts
of it) can interact with thermal baths at different temperatures. 

Until Sec. \ref{sec:Tsalis and Renyi}, our main object of interest
is the $F$-shifted $\alpha$ energy moment
\begin{align}
\mathcal{H}_{\alpha} & =\left\langle [H(t)-F(t)]^{\alpha}\right\rangle =\sum_{j}p_{j}[E_{j}(t)-F(t)]^{\alpha},\label{eq: Halpha def}\\
F(t) & =T(t)\ln\sum_{j}\exp[-E_{j}(t)/T(t)]
\end{align}
where $\alpha$ is a \textit{real} and \textit{positive} number ($\alpha$
is not necessarily an integer). This quantity is an observable that
can be evaluated from energy measurements. It contains information
on higher moments of the energy distribution. The appearance of the
instantaneous free energy $F(t)$ is interesting. First, it makes
$H-F$ a positive operator so any $\alpha\ge0$ power of $H-F$ is
well defined. Second, it makes $(H-F)^{\alpha}$ invariant to uniform
shifts of all the levels by a constant (in contrast to a regular moment
$\left\langle H^{\alpha}\right\rangle $). The variance, for example,
is also shift-invariant but the subtraction of the average makes the
variance a nonlinear function of $p_{j}$ which significantly complicates
the analysis. Nonetheless, in a certain class of cases $\mc H_{2}$
will be equal to to the energy variance. The appearance of $F(t)$
in (\ref{eq: Halpha def}) follows from the derivation of the GCI
as explained below.

Now that we have an energy related quantity, it is possible define
its flows in the same way it is done for the average energy (e.g.
\cite{anders2013thermodynamics})
\begin{eqnarray}
\mc W_{\alpha} & \triangleq & \int_{t_{i}}^{t_{f}}dt\sum_{j}p_{j}\frac{d}{dt}[(E_{j}-F)^{\alpha}],\label{eq: Wa def}\\
\mc Q_{\alpha} & \triangleq & \int_{t_{i}}^{t_{f}}dt\sum_{j}\frac{dp_{j}}{dt}(E_{j}-F)^{\alpha}.\label{eq: Qa def}
\end{eqnarray}
In the present paper subscript $'f'$ stands for 'final' and $'i'$
stands for 'initial' (to prevent confusion '$i$' will not be used
as a summation index). Just like in the $\alpha=1$ case, the logic
behind these definitions is that if the levels are fixed in time,
the changes in energy must be due to heat exchange with the environment.
If the populations are fixed the change in energy must be work related.
From (\ref{eq: Wa def}) and (\ref{eq: Qa def}) we get
\begin{equation}
\Delta\mc H_{\alpha}=\mc Q_{\alpha}+\mc W_{\alpha}.\label{eq: H alpha conserv}
\end{equation}
We first consider two elementary thermodynamic primitives: 1) Isochores:
the system is coupled to a bath and the energy levels $E_{j}$ do
not change in time. 2) Adiabats: the system is not connected to a
bath, and the energy levels change in time (energy populations are
fixed in time). Note that adiabats need not be slow. The use of this
term refers to an 'adiabatic' process in macroscopic classical thermodynamics,
where the system is isolated from the environment, and consequently
the entropy of the system does not change in time. Isochores involve
only heat, while adiabats involve only work. Other processes such
as isotherms can be constructed by concatenating these two primitives
\cite{anders2013thermodynamics}. We now proceed to the derivation
of an important family of the GCI the $\alpha$CI. 

Let us first consider a basic isochore thermalization process: the
system is connected to a single bath with inverse temperature $\beta$
and the levels of the system do not change in time. In Appendix I
we show that for isochores
\begin{equation}
\Delta\mc S_{\alpha}-\frac{\mc Q_{\alpha}}{T^{\alpha}}=D_{\alpha}^{B}(\vec{p}_{i},\vec{p}_{\beta})-D_{\alpha}^{B}(\vec{p}_{f},\vec{p}_{\beta}).\label{eq: Bregman isochore}
\end{equation}
On the left hand side (LHS) we introduce the $\alpha$ information
function $\mc S_{\alpha}$ which is defined as
\begin{align}
\mc S_{\alpha} & = & \sum_{j}\intop_{c_{0}}^{p_{j}}(-\ln x)^{\alpha}dx=[\sum_{j}\Gamma(\alpha+1,-\ln p_{j})]\nonumber \\
 &  & -\Gamma(\alpha+1,0)\label{eq: S def}
\end{align}
Where $\Gamma$ is the indefinite Gamma function. The lower limit
of the integral adds a constant term to $\mc S_{\alpha}$. For convenience,
$c_{0}$ was chosen so that $\mc S_{\alpha}=0$ for deterministic
states (one probability is equal to one and the rest are equal to
zero). $\mc S_{1}$ is the standard Shannon entropy $S=-\sum p_{j}\ln p_{j}$.
In Appendix V we comment on the physicality of $\mc S_{\alpha}$ compared
to the standard Shannon entropy. In short, we argue that at least
in our context $\mc S_{\alpha}$ and $S$ have the same functionality:
both are used to put restrictions on energy changes created by a thermal
bath. The relation of $\mc S_{\alpha}$ to information is discussed
after stating the GCI.

On the right hand side (RHS) of (\ref{eq: Bregman isochore}) we have
the Bregman divergence of the initial state of the system $\vec{p}_{i}$
and the final state of the system $\vec{p}_{f}$ with respect to the
thermal state $\vec{p}_{\beta}$. The definition of the Bregman divergence
$D_{\alpha}^{B}(\vec{p}_{2},\vec{p}_{1})$ is \cite{Bregman1967}
\begin{equation}
D_{\alpha}^{B}(\vec{p}_{2},\vec{p}_{1})\doteq\mc S_{\alpha}(\vec{p}_{1})-\mc S_{\alpha}(\vec{p}_{2})+(\vec{p}_{2}-\vec{p}_{1})\cdot\nabla\mc S_{\alpha}(p_{1})\label{eq: Bregman def}
\end{equation}
This divergence and its appealing geometric interpretation are described
in Appendix I. For now, it suffices to know only several of its key
features. First, mathematically, it is a divergence so it satisfies
$D_{\alpha}^{B}(\vec{p}_{2},\vec{p}_{1})\ge0$ and $D_{\alpha}^{B}(\vec{p}_{2},\vec{p}_{1})=0\:\Leftrightarrow\vec{p}_{2}=\vec{p}_{1}$.
Second, $D_{\alpha}^{B}(\vec{p}_{2},\vec{p}_{1})$ is convex in the
first argument $\vec{p}_{2}$ (see Appendix I). Third, for $\alpha=1$,
$D_{1}^{B}$ is the Kullback-Leibler Divergence (relative entropy)
$D_{KL}(\vec{p}_{2},\vec{p}_{1})=\sum_{j}p_{2,j}\ln(p_{2,j}/p_{1,j})$.

Equation (\ref{eq: Bregman isochore}) is an identity valid for isochore.
To make use of it we need to make some physical statements on one
of the sides of side of the identity. A thermal interaction is a map
$\mc M_{\beta}(\vec{p}_{i})=\vec{p}_{f}$ with the thermal state as
a fixed point $\mc M_{\beta}(\vec{p}_{\beta})=\vec{p}_{\beta}$. The
RHS of (\ref{eq: Bregman isochore}) is a measure of contractiveness
of the map. $D_{\alpha}^{B}(\vec{p},\vec{p}_{\beta})$ can be regarded
as a proximity measure (a divergence, not necessarily a distance)
between the state $\vec{p}$ and $\vec{p}_{\beta}$. Thus, a positive
value in the RHS of (\ref{eq: Bregman isochore}) implies $\mc M_{\beta}$
is contractive with respect to $\vec{p}$ (under the $D_{\alpha}^{B}$
measure), i.e. $\vec{p}_{f}$ is closer than $\vec{p}_{i}$ to the
fixed point $\vec{p}_{\beta}$ (in terms of $D_{\alpha}^{B}$). 

While the LHS is the content of the physical law, the RHS sets its
regime of validity. Consider the case of full thermalization where
$\vec{p}_{f}=\vec{p}_{\beta}$. Since $D_{\alpha}^{B}(\vec{p}_{f}=\vec{p}_{\beta},\vec{p}_{\beta})=0$
it follows from (\ref{eq: Bregman isochore}) that $\Delta\mc S_{\alpha}-\frac{\mc Q_{\alpha}}{T^{\alpha}}\ge0$.
In Sec. \ref{subsec: Functional-definition} we show that in addition
to the contractiveness interpretation of the RHS, it also has an appealing
thermodynamic interpretation related to reversible processes and maximal
work extraction.

In Sec. \ref{subsec: Regime-of-validity} we discuss cases where the
RHS of (\ref{eq: Bregman isochore}) is guaranteed to be positive,
but for now let us assume it is positive for isochores and see how
various thermodynamic results emerge. Consider a thermodynamic protocol
composed of an infinitesimal concatenation of isochores and adiabats.
It is assumed that the isochores are short enough so that the temperature
of the bath in each isochore is fixed. Hence, isochore $'l'$ satisfies
$\delta\mc S_{\alpha}^{(l)}-\frac{1}{T_{(l)}{}^{\alpha}}\mc{\delta Q}_{\alpha}^{(l)}\ge0$.
The adiabats that connect the isochores carry zero $\alpha$ heat
$\delta\mc Q_{\alpha}=0$ and they do not change the entropy $\delta\mc S_{\alpha}=0$.
Summing the infinitesimal contributions of both isochores and adiabats
we get our first main finding: the $\alpha$CI family of the GCI's
\begin{equation}
\Delta\mc S_{\alpha}-\int\frac{\frac{d}{dt}\mc Q_{\alpha}}{T(t)^{\alpha}}dt\ge0\label{eq: CI ge 0}
\end{equation}
See (\ref{eq: gen CI}) for a more general form with simultaneous
connection to multiple baths. In the study of heat machines the periodic
form of the second law $-\oint\frac{\mc{\delta Q}(t)}{T(t)}\ge0$
is highly useful. For a periodic protocol, as in heat engines and
refrigerators, the system reaches a cyclic operation $\vec{p}(t+\tau_{cyc})=\vec{p}(t)$
where $\tau_{cyc}$ is the cycle time. Since in this case $\Delta\mc S_{\alpha}=\mc S_{\alpha}[\vec{p}(t+\tau_{cyc})]-\mc S_{\alpha}[\vec{p}(t)]=0$
we get the analogue of the periodic CI
\begin{equation}
-\oint\frac{\mc{\delta Q}_{\alpha}(t)}{T^{\alpha}(t)}\ge0\label{eq: periodic form}
\end{equation}
where $\delta\mc Q_{\alpha}$ is the $\alpha$ heat transferred during
a short time $dt$. The advantage of this form is that it is completely
free of $\mc S_{\alpha}$.

In isotherms (IST), the system is always in the Gibbs state $p_{\beta}=p_{\beta(t)}[H(t)]=e^{-\beta(t)[H(t)-F(t)]}$
even though $\beta$ and/or the energy levels can vary in time). As
shown in the end of Appendix I, by considering an infinitesimal concatenation
of isochores and adiabats one can show that 
\begin{equation}
\Delta\mc S_{\alpha}^{IST}-\int\delta\mc Q_{\alpha}^{IST}/T(t)^{\alpha}=0\label{eq: isotherm}
\end{equation}
There is an alternative and simple way to obtain relation (\ref{eq: isotherm})
for isotherms. Consider a protocol where the system Hamiltonian is
changed in time $H=H(t)$, but the system is always in thermal equilibrium
$p_{\beta(t)}(H(t))=e^{-\beta(t)[H(t)-F(t)]}$ (for this the change
in H(t) should be much slower than the thermalization time). Using
(\ref{eq: Qa def}) and (\ref{eq: S def}) for isotherms $p=p_{\beta}(t)$
we find
\begin{align}
\intop\frac{\delta\mc Q_{\alpha}^{IST}}{T(t)^{\alpha}} & =\sum_{j}\intop_{p_{i}}^{p_{fin}}\frac{(H-F)_{j}^{\alpha}}{T(t)^{\alpha}}dp_{,j}\nonumber \\
 & =\sum_{j}\intop_{p_{i}}^{p_{fin}}(-\ln p_{\beta,j})^{\alpha}dp_{\beta,j}\nonumber \\
 & \equiv S_{\alpha}(\vec{p}_{\beta,fin})-S_{\alpha}(\vec{p}_{\beta,i})\label{eq: Q iso proof}
\end{align}
Therefore we obtain the equality of the GCI $\Delta\mc S_{\alpha}^{IST}=\intop\beta(t)^{\alpha}\delta\mc Q_{\alpha}^{IST}$.

Reversible processes consist of (non-infinitesimal) sequences of isotherms
and adiabats (for adiabats $\Delta\mc S_{\alpha}=\mc Q_{\alpha}=0$)
so we can write
\begin{equation}
\Delta\mc S_{\alpha}-\int\frac{\delta\mc Q_{\alpha}^{R}}{T(t)^{\alpha}}=0.\label{eq: CI rev}
\end{equation}
where $\mc Q_{\alpha}^{R}$ is the heat absorbed in a reversible process.
We now wish to clarify the difference between (\ref{eq: isotherm})
and (\ref{eq: CI rev}). Any isotherm must start and end in thermal
equilibrium. Thus, the endpoints of adiabats between isotherms are
fully determined. However, an adiabat at the end of the protocol need
not end at a thermal state. Consequently, a reversible process may
involve a final state that is very different from a thermal state
(a similar argument can be applied for the initial state). Nonetheless,
(\ref{eq: CI rev}) states that the equality in the GCI is valid also
for reversible processes that start or end out of equilibrium.

Finally, we conclude that \textit{the GCI's are not just inequalities;
they are inequalities that are saturated for reversible processes}.
In perfect analogy to the standard second law, if a reversible process
is given the CI implies that all possible irreversible process with
the same endpoint, will be less optimal (e.g. produce less work and
consume more heat). This is discussed in detail in Sec. \ref{subsec: Single-bath-forms and state prep}. 

\subsection{Information in the GCI and extensivity}

By virtue of the GCI $\mc S_{\alpha}$ is the information conjugated
to $\mc Q{}_{\alpha}$ heat. The reasons for associating $\mc S_{\alpha}$
with information are the following: 1) $\mc S_{\alpha}$ of a pure
(deterministic) state is zero. 2) $\mc S_{\alpha}$ is symmetric.
It is invariant to rearrangement (permutation) of the probabilities.
3) It increases under doubly stochastic transformations, and it obtains
a maximal value for the uniform distribution. The third property follows
from the fact that $\mc S_{\alpha}$ is Schur concave. The Schur concavity
is a built-in feature of the Bregman formalism described in Appendix
I. The reason for expecting this property is that doubly stochastic
transformations are mixtures of permutations, and they smear out the
probability distribution and make it more random. 

Note that we have not imposed further requirements on the information
measure such as extensivity. $\mc Q{}_{\alpha}$ is in general not
an extensive quantity so the information conjugated to it need not
be extensive as well. 

\subsection{Single bath forms and reversible state preparation \label{subsec: Single-bath-forms and state prep}}

In this section we study the case where a single bath with a fixed
temperature $T$ is available to interact with the system (in contrast
to the more general $T(t)$ used until now). From (\ref{eq: CI ge 0})
and (\ref{eq: CI rev}) we conclude that in the validity regime of
the CI, any process that includes adiabats, isotherms, and isochores
satisfies

\begin{eqnarray}
\mc Q_{\alpha} & \le & \mc Q_{\alpha}^{R},\label{eq: Q QR}\\
\mc W_{\alpha} & \ge & \mc W_{\alpha}^{R},\label{eq: W WR}
\end{eqnarray}
where $\mc Q_{\alpha}^{R},\mc W_{\alpha}^{R}$ are the $\alpha$ reversible
heat, and $\alpha$ reversible work gained by going from $\{\vec{p}_{i},H_{i}\}$
to $\{\vec{p}_{f},H_{f}\}$ in a reversible protocol. $\mc Q_{\alpha},\mc W_{\alpha}$
are the heat and work gained in an \textit{irreversible} process between
the same $\{\vec{p},H\}$ endpoints. Equation  (\ref{eq: W WR}) is
obtained from (\ref{eq: Q QR}) and (\ref{eq: H alpha conserv}).

In analogy to standard thermodynamics, the reversible $\alpha$ work
that can be extracted by going from $\{\vec{p}_{i},H_{i}\}$ to $\{\vec{p}_{f},H_{f}\}$,
takes the form
\begin{eqnarray}
\mc W^{R} & = & \Delta\mc F_{\alpha}-T^{\alpha}D_{\alpha}^{B}(\vec{p}_{i},\vec{p}_{\beta,i})+T^{\alpha}D_{\alpha}^{B}(\vec{p}_{f},\vec{p}_{\beta,f}),\nonumber \\
\label{eq: WR berg}\\
\vec{p}_{\beta,i(f)} & = & exp[-\beta(H_{i(f)}-F_{i(f)})],
\end{eqnarray}
where $\mc F_{\alpha}=\mc H_{\alpha}(\vec{p}_{\beta})-T^{\alpha}\mc S_{\alpha}(\vec{p}_{\beta})$
is the $\alpha$ order (equilibrium) free energy. Equation (\ref{eq: WR berg})
is proven in Appendix II. Note that one can also define a quantity
$\tilde{\mc F}_{\alpha}(\vec{p})=\mc{H_{\alpha}}(\vec{p})-T^{\alpha}\mc S_{\alpha}(\vec{p})$
that is the $\alpha$CI analogue of the non-equilibrium free energy
\cite{CrooksThemoPredNonEf,Esposito2011EPL2Law} (for equilibrium
states $\tilde{\mc F}_{\alpha}(\vec{p}_{\beta})=\mc F_{\alpha}$).
With this definition the $\mc W_{\alpha}^{R}$ is given by $\mc W_{\alpha}^{R}=\Delta\tilde{\mc F}_{\alpha}$.

We point out that Landauer erasure \cite{reeb2014improved} is a special
case of thermal state preparation. Thus, the reversible limit (\ref{eq: Q QR})-(\ref{eq: WR berg})
bounds the changes in $\alpha$ moments in erasure scenarios as well.

From (\ref{eq: Q QR}) and (\ref{eq: W WR}), we deduct the single
bath cyclic formulation of the second law $\oint d\mc W_{\alpha}\ge0$
: it is impossible to extract $\alpha$ work in a periodic process
with a single bath. Similarly, $\oint d\mc Q_{\alpha}\le0$ implies
that an $\alpha$ heat cannot be extracted from a single bath in a
periodic process (the changes in the bath are studied in Appendix
III). 

\subsection{Regime of validity\label{subsec: Regime-of-validity} }

From (\ref{eq: Bregman isochore}) the $\alpha$CI's (\ref{eq: CI ge 0})
validity condition for a single bath isochore is
\begin{equation}
D_{\alpha}^{B}(\vec{p}_{i},\vec{p}_{\beta})-D_{\alpha}^{B}(\vec{p}_{f},\vec{p}_{\beta})\ge0.\label{eq: validity cond}
\end{equation}
Next, it is shown that this validity condition is guaranteed to hold
in the following cases:
\begin{itemize}
\item Strong thermalization: $\vec{p}_{f}$ is equal or sufficiently close
to $\vec{p}_{\beta}$. 
\item Uniform thermalization map $\vec{p}_{f}=(1-y)\vec{p}_{i}+y\vec{p}_{\beta}$
where $0\le y\le1$.
\item Two-level system
\item Isotherms
\item Adiabats
\end{itemize}
The first regime follows from the fact that $D_{\alpha}^{B}(\vec{p}_{f}\to\vec{p}_{\beta},\vec{p}_{\beta})\to0$
so that the RHS of (\ref{eq: Bregman isochore}) is positive. This
is a very important regime as it can take place in different thermalization
mechanisms and \textit{also in the presence of initial correlation
with the bath}. Usually for the $\alpha=1$ CI it is assumed that
the system and bath \cite{Sagawa2012second,PeresBook} are initially
uncorrelated. However, this is not a necessary condition when the
final state is thermal or very close to it. Thus, the first validity
regime is indifferent to initial system-bath correlations for any
$\alpha>0$. 

The second regime follows from the fact that the Bregman divergence
(\ref{eq: Bregman def}) is convex in its first argument \footnote{Follow immediately from the convexity of $-\mc S$ and the fact that
a linear term does not affect concavity.}. The third regime holds since the thermalization of a two-level system
is always uniform so the two-level case is always contained in the
second regime. For isotherms each of the divergence terms in (\ref{eq: validity cond})
is zero so the $\alpha$CI holds as an equality. For adiabats $\vec{p}_{f}=\vec{p}_{i}$,
and once again we get zeros in (\ref{eq: validity cond}).

Potentially, the regime of validity is much larger than outlined above.
Numerical studies showed that significant deviations from the uniform
thermalization map still satisfy (\ref{eq: validity cond}). This
is a subject for further research. Nevertheless, the examples given
later demonstrate that the above regimes are already sufficient for
showing that the various $\alpha$CI bring new insights to thermodynamic
scenarios of microscopic systems.

For $\alpha=1$, the $\alpha$CI reduce to the CI and the $D^{B}$
in the validity condition in (\ref{eq: validity cond}) reduces to
the relative entropy. When the thermalization can be described by
a completely positive trace preserving map (CPTP) with a Gibbs state
as a fixed point, the relative entropy of a state with respect to
a fixed point of the map is always decreasing (including quantum dynamics).
This means that condition (\ref{eq: validity cond}) for $\alpha=1$
is satisfied for such thermalization maps. CPTP maps arise naturally
when an initially uncorrelated system and a bath interact via a unitary
operation. The thermal operations in thermodynamic resource theory
are an example of such operations. From this we conclude that for
$\alpha=1$ our derivation has the same validity regime as that obtained
from other derivations based on CPTP maps \cite{PeresBook,Sagawa2012second,Esposito2011EPL2Law,spohn78,breuer,alicki79}.

For $\alpha\neq1$ condition (\ref{eq: validity cond}) may not be
satisfied for CPTP. For example, consider a three-level system where
all levels are initially populated. If only levels two and three are
coupled to a bath then in general (\ref{eq: validity cond}) may not
hold for $\alpha\neq1$ (even though the ratio of $p_{3}/p_{2}$ gets
closer to the Gibbs factor $exp[-\beta(E_{3}-E_{2})]$).

A smaller regime of validity is not always a disadvantage. The invalidity
of one of the $\alpha$CI can give us information on the thermalization
process under progress. For example, if in a periodic system (\ref{eq: periodic form})
is not satisfied we can rule out any of the thermalization scenarios
described in the bullets above. Moreover, a regime of invalidity might
be useful for some purposes since it is less constrained. Applying
such ideas for $\alpha\neq1$ in heat machines is outside the scope
of this paper.

\subsection{Allowed operations}

Analogously to thermodynamic resource theory \cite{Goold2015review,horodecki2013fundamental,GourRTreview},
the regime of validity can be formulated in terms of allowed operations.
Instead of giving a validity condition it is possible to restrict
the set of allowed physical processes. Assuming that the system starts
in valid initial condition, any allowed operation will keep the system
in the validity regime. For example, in a restricted set of operations
that includes only full thermalization, uniform thermalization, and
adiabats, the generalized Clausius inequalities hold. 

\subsection{Functional definition of a bath and reversible work availability\label{subsec: Functional-definition}}

Equation (\ref{eq: Bregman isochore}) is the quintessence of the
second law (CI more accurately): both the standard CI and the generalizations
that are considered here. Thermodynamics describes interactions with
baths. What is a bath, then? There are two main answers and both are
useful. One approach describes the bath and its physical properties
such as temperature, correlation function, heat capacity etc. As a
second approach we suggest the functional definition of a bath. The
ideal operation of a bath would be to take any initial state of a
system and change some of its properties (e.g. energy moments, or
other observables) into predefined values that are independent of
the initial state. For example, a thermal bath takes any initial state
to a Gibbs state $\vec{p}_{\beta}$ with temperature $1/\beta$. The
Gibbs state is the fixed point of the map the bath induces on the
system. 

In practice the bath is not connected to the system for an infinite
amount of time so the process may not be completed. Moreover, if the
bath is small compared to the system, it may not have enough energy
to complete the thermalization of the system (even though the Gibbs
state is a fixed point of this interaction). Nonetheless, it is expected
that the final state in these scenarios will be ``closer'' to the
thermal state compared to the initial state. What is the measure of
proximity we should use in order to make sure we have valid thermodynamic
laws e.g. (\ref{eq: CI ge 0})? This is exactly the Bregman divergence
difference (\ref{eq: validity cond}).

Equation (\ref{eq: Bregman isochore}) tells us that if the RHS is
negative it simply indicates that the device we used for a bath has
failed to operate as a bath since it did not decrease the proximity
measure of interest. 

A ``good bath'' is one that satisfies (\ref{eq: validity cond})
for any $\vec{p}_{i}$, however for specific applications such as
heat machines it may be sufficient that this condition holds only
for the $\vec{p}_{i}$ and $\vec{p}_{f}$ of interest (e.g. those
that appear in cyclic operation). 

Condition (\ref{eq: validity cond}) has a very appealing thermodynamic
interpretation. From (\ref{eq: W WR}) and (\ref{eq: WR berg}) we
see that $T^{\alpha}D_{\alpha}^{B}(\vec{p}_{i},\vec{p}_{\beta})$
expresses the maximal $\alpha$ work that can be extracted by going
from $\vec{p}_{i}$ to $\vec{p}_{\beta}$ when the initial Hamiltonian
and final Hamiltonian are equal (but may be changed in the middle)
so that $\Delta\mc F_{\alpha}=0$. Let us use the term ``available
$\alpha$ work'' $\mc A_{\alpha}(\vec{p}_{i})\triangleq T^{\alpha}D_{\alpha}^{B}(\vec{p}_{i},\vec{p}_{\beta})$.
The quantity $\mc A_{\alpha}$ is analogous to the ergotropy \cite{AllahverdyanErgotropy}
in closed quantum systems. The only difference is that ergotropy quantifies
the work that can be extracted using only unitaries, while here a
bath can be connected as well. 

In general, when thermalizing we expect that $\mc A_{\alpha}$ will
decrease. This means that the thermal bath makes the system less ``thermally
active''. In full thermalization the state becomes $\vec{p}_{\beta}$
which is ``thermally passive'' ($\mc A_{\alpha}=0$).

\subsection{Intermediate summary}

The underlying principles of the thermodynamic generalized Clausius
inequalities as studied so far in this paper can be summarized as
follows:
\begin{itemize}
\item Changes in $\alpha$ order energy moments are associated with changes
in a corresponding information measure $\mc S_{\alpha}$. The two
are related by $\alpha$ order Clausius inequality (\ref{eq: CI ge 0})
and (\ref{eq: gen CI}). The GCI's establish new energy-information
relations.
\item The generalized Clausius inequalities saturate and become equalities
in reversible thermodynamic processes. 
\item Different energy moments are associated with different divergences
that constitute a validity criterion for the functionality of the
bath. When the bath brings the system closer to the thermal state
according to the $\alpha$ divergence measure, the generalized $\alpha$
order Clausius inequality holds.
\item Several important $\alpha$CI validity regimes have been identified
but it is important to explore and find additional regimes. 
\end{itemize}

\section{Additional second laws based on Rényi entropy and $\alpha$ impurity\label{sec:Tsalis and Renyi}}

In the $\alpha$CI we started with some energy moments of interest
(\ref{eq: Halpha def}) and found the corresponding information measure
that is related to it via the $\alpha$CI. In this section we go in
the other direction: we pick information quantities of interest and
find the moments that are related to them via the GCI. 

As shown in Appendix I, $\partial_{p}\mc S(p)$ can be any monotonically
decreasing function in $p\in[0,1]$, and lead to a generalized Clausius
inequality. Alternatively, $\mc S$ can be any concave and differentiable
function in the regime $p\in[0,1]$. Let us first choose $\mc S$
to be the $\tilde{\alpha}$ impurity \cite{havrda1967EarlyTsallisDef,vajda1968EarlyTsallisDef,daroczy1970EarlyTsallisDef,TsalisOriginal,TsalisStatMech}
$\mc S_{\tilde{\alpha}}(\vec{p})=\frac{1}{\tilde{\alpha}-1}(1-\sum p_{j}^{\tilde{\alpha}})$
($\tilde{\alpha}\ge0$ can be a fraction). This quantity is often
called ``Tsallis entropy'' \cite{TsalisOriginal,TsalisStatMech},
but to prevent unnecessary technical confusions we use here the term
``$\tilde{\alpha}$ impurity''. \footnote{Purity is defined $tr[\rho^{2}]$ and we call one minus the purity
the ``impurity'' of $\rho$. The $\tilde{\alpha}$ impurity is best
known as Tsallis entropy. However, the physical context of Tsallis
entropy is associated with non thermal Tsallis distribution. Since
our reservoirs are always thermal we use a different name to prevent
possible confusion.} . For $\tilde{\alpha}\ge0$, $\partial_{p}\mc S_{\tilde{\alpha}}$
is a decreasing function and we can use our GCI formalism and get
the analogue of (\ref{eq: Bregman isochore}) for the $\alpha$ impurity
\begin{align}
\Delta\mc S_{\tilde{\alpha}}^{B}+\frac{\tilde{\alpha}}{\tilde{\alpha}-1}\Delta\left\langle e^{-(\tilde{\alpha}-1)\beta(H-F)}\right\rangle  & =D_{\tilde{\alpha}}^{B}(\vec{p}_{i},\vec{p}_{\beta})\nonumber \\
 & -D_{\tilde{\alpha}}^{B}(\vec{p}_{f},\vec{p}_{\beta})\label{eq: CI Tsalis}
\end{align}
The GCI (\ref{eq: CI Tsalis}) is written here for simplicity just
for isochores. From the time derivative of the observable $+\frac{\tilde{\alpha}}{\tilde{\alpha}-1}\left\langle e^{-(\tilde{\alpha}-1)\beta(H-F)}\right\rangle $
one can define $\tilde{\alpha}$ heat and $\tilde{\alpha}$ work and
get an $\tilde{\alpha}$CI valid for isochores, isotherms adiabats,
and their combination (as done in Sec. 2)
\begin{align}
\Delta\mc S_{\tilde{\alpha}}^{B}-\intop\delta Q_{\tilde{\alpha}} & \ge0,\label{eq: impurity gCI}\\
\delta Q_{\tilde{\alpha}} & =\frac{\tilde{\alpha}}{\tilde{\alpha}-1}\sum_{j}\delta p_{j}e^{-(\tilde{\alpha}-1)\beta(E_{j}-F)}.
\end{align}
 As in the previous generalized CI, when setting $\tilde{\alpha}=1$
in (\ref{eq: CI Tsalis}) it reduces to the standard Clausius inequality.
Another important case is $\tilde{\alpha}=2$ where $\mc S_{\tilde{\alpha}}^{B}$
is equal to minus the purity of the state (plus a constant), and the
Bregman divergence is the standard Euclidean distance squared $D_{\tilde{\alpha}}^{B}(\vec{p}_{i},\vec{p}_{\beta})=\nrm{\vec{p}_{i}-\vec{p}_{\beta}}_{2}^{2}$.
While in general the Bregman divergence is not a distance, for $\tilde{\alpha}=2$
it is. The regime of validity is given by the positivity of the RHS
of (\ref{eq: CI Tsalis}). The regime of guaranteed validity is at
least as large as that given in Sec. \ref{subsec: Regime-of-validity}.
This time the information we obtain is on $\left\langle e^{-(\tilde{\alpha}-1)\beta(H-F)}\right\rangle =\sum p_{j}e^{-(\tilde{\alpha}-1)\beta(E_{j}-F)}$.
The information on higher order moments of the distribution is wrapped
in an exponential form. In fact, this exponential form is the moment
generating function of the distribution. Using the Markov inequality
it is possible to learn about the tail of the distribution $P(e^{-(\tilde{\alpha}-1)\beta(E_{j}-F)}\ge\xi)\ge\frac{\left\langle e^{-(\tilde{\alpha}-1)\beta(H-F)}\right\rangle }{\xi}$.
Another advantage of this form is that the free energy can easily
be pulled out and be replaced by a different constant $E_{ref}$ :
\begin{align}
\left\langle e^{-(\tilde{\alpha}-1)\beta(H-F)}\right\rangle  & =\nonumber \\
 & e^{-(\tilde{\alpha}-1)\beta(E_{ref}-F)}\left\langle e^{-(\tilde{\alpha}-1)\beta(H-E_{ref})}\right\rangle .\label{eq: Factor out F}
\end{align}
For example, $E_{ref}$ can be the ground state energy or the average
energy. For $E_{ref}=0$ (\ref{eq: Factor out F}) shows that $F$
is a factor that can be pulled out from the expectation value (in
contrast to $\mc H_{\alpha}$). This is useful when considering the
interaction with the bath (see Appendix III). 

Here as well, there is a periodic form of the $\tilde{\alpha}$ second
law
\begin{equation}
\frac{1}{\tilde{\alpha}-1}\oint_{k}\sum_{j}dp_{j}e^{-(\tilde{\alpha}-1)\beta_{k}(t)[E_{k,j}(t)-F_{k}(t)]}\ge0\label{eq: period Tsalis}
\end{equation}
The $\oint_{k}$ symbol stands for summation over connection to different
baths $\beta_{k}$ during a cycle. 

The GCI can be applied to the Rényi entropy $R_{\bar{\alpha}}(\vec{p})=\frac{1}{1-\bar{\alpha}}\ln\sum p_{j}^{\bar{\alpha}}$
as well. This time $\bar{\alpha}$ is limited to $0\le\bar{\alpha}\le1$
where the Rényi entropy is concave. For $\bar{\alpha}=1$ the Rényi
entropy reduces to the Shannon entropy. The $\bar{\alpha}$CI is given
by (\ref{eq: CI Tsalis}) with $\mc S_{\tilde{\alpha}}^{B}\to R_{\bar{\alpha}}$
and $\delta\mc Q_{\tilde{\alpha}}\to\delta\mc Q_{\bar{\alpha}}$

\begin{equation}
\delta Q_{\bar{\alpha}}=\frac{\bar{\alpha}}{1-\bar{\alpha}}\frac{\sum_{j}\delta p_{j}e^{-\beta(\bar{\alpha}-1)(E_{j}-F)}}{\sum_{j}(p_{\beta,j})^{\bar{\alpha}}}.\label{eq: Renyi CI}
\end{equation}
The validity regime is at least as large as described in Sec. \ref{subsec: Regime-of-validity}.
In the $\bar{\alpha}$CI analogue of (\ref{eq: Bregman isochore})
the divergence is given by the Bregman divergence with $R_{\bar{\alpha}}$
instead of $\mc S_{\alpha}$. This divergence is \textit{not} the
Rényi divergence that is used in thermodynamic resource theory \cite{Goold2015review,horodecki2013fundamental,GourRTreview}. 

\subsection{High temperature limit\label{subsec: High-temperature-limit}}

In this section we study the high temperature limit of the $\tilde{\alpha}$
impurity GCI ($\tilde{\alpha}$CI). First we consider periodic operation
(\ref{eq: period Tsalis}) where $\mc S$ does not appear. For any
$\beta$ that is small enough, the Taylor expansion of the exponent
in (\ref{eq: period Tsalis}) leads to the standard periodic form
of the second law $-\oint\frac{dQ}{T}\ge0$. This is interesting until
now we to obtain the CI from a GCI we took the limit $\alpha\to1,\tilde{\alpha}\to1,$
etc. However here we see that the periodic form emerges for high temperatures
even when $\tilde{\alpha}\neq1$. To be more precise the condition
for being hot enough emerges from the expression.
\begin{align}
\left\langle e^{-\beta(\tilde{\alpha}-1)(H-F)}\right\rangle  & =e^{-\beta(\tilde{\alpha}-1)(\frac{E_{max}+E_{min}}{2}-F)}\nonumber \\
 & \left\langle e^{-\beta(\tilde{\alpha}-1)(H-\frac{E_{max}+E_{min}}{2})}\right\rangle 
\end{align}
In order to approximate the exponent in the triangular brackets with
a linear term in $H$ the temperature must satisfy the condition
\begin{equation}
T\gg\left|\tilde{\alpha}-1\right|(E_{max}-E_{min})/2.\label{eq: refined high T cond}
\end{equation}
When this condition holds (\ref{eq: period Tsalis}) reduces to the
periodic CI $-\oint\frac{\mc{\delta Q}(t)}{T(t)}\ge0$. A more interesting
result appears in the non-periodic form of the $\tilde{\alpha}$CI.
Using (\ref{eq: refined high T cond}) in (\ref{eq: impurity gCI})
we get
\begin{equation}
\Delta\mc S_{\tilde{\alpha}}-\tilde{\alpha}e^{+\beta(\tilde{\alpha}-1)(\frac{E_{max}+E_{min}}{2}-F)}Q/T\ge0.\label{eq: tsalis high T isochore}
\end{equation}

Equation (\ref{eq: tsalis high T isochore}) implies that on top of
the Shannon entropy there are \textit{different information measures}
that put additional constraints on the \textit{standard heat exchange}
with the bath. While (\ref{eq: tsalis high T isochore}) offers no
advantage for reversible processes, for irreversible processes it
can provide a tighter bound on the heat. An example for the superiority
of (\ref{eq: tsalis high T isochore}) over (\ref{eq: CI alpha=00003D1})
is given in Sec. \ref{sec: Examples}.

This finding paves the way for studying information measures that
provide tighter bounds (compared to the second law) on the heat in
irreversible processes in various limits (e.g. in cold temperatures).
Although we have focused on the $\tilde{\alpha}$ impurity GCI, the
high temperature limit can be studied for the Rényi entropy as well.

\section{Quantum generalized CI and thermodynamic coherence measures\label{sec: Quantum gCI}}

Let $g(x)$ be an analytic and concave function of $x$. We denote
its integral by $G(x)=\intop_{x_{0}}^{x}g(x')dx'$ where $x_{0}$
will be chosen later. The Bregman matrix divergence \cite{FNielsen2013miningBregmanMatrix,molnar2016mapsBregmanMatrix}
is \footnote{We use the Bregman divergence only for density matrices. Hence, conjugation
and transposition are not needed.}
\begin{align}
D_{g}^{B}(\rho_{2},\rho_{1}) & =\mc S_{g}(\rho_{1})-\mc S_{g}(\rho_{2})+tr[(\rho_{2}-\rho_{1})g(\rho_{1})]\label{eq: QBreg}\\
\mc S_{g}(\rho_{1}) & =tr[G(\rho)]\label{eq: S_g def}
\end{align}
Repeating the stochastic derivation carried out in Sec. II and in
Appendix I, we obtain the quantum generalized CI (QGCI):
\begin{align}
\Delta\mc S_{g}-\intop\frac{\delta\mc Q_{\alpha}}{T^{\alpha}} & \ge0,\label{eq: QalphaCI}\\
\delta\mc Q_{\alpha} & =tr[\delta\rho(H-F)^{\alpha}].
\end{align}
We wrote down the quantum form of the stochastic $\alpha$CI (\ref{eq: CI ge 0}),
but it is equally possible to write the quantum analogue of the other
forms (e.g. the Rényi form (\ref{eq: Renyi CI})). To obtain the $\alpha$
form we have set $g(x)=(-\ln x)^{\alpha}$. This time, $\mc S_{g}$
(\ref{eq: S_g def}) is defined for density matrices with coherence.
$\mc S_{g}$ has several properties we expect from information measures:
1) it is unitarily invariant (extension of the symmetry property in
the stochastic case) 2) it increases under doubly stochastic maps
3) $min(\mc S_{g})=0$ is obtained for pure states $\ketbra{\psi}{\psi}$
($x_{0}$ can always be chosen so that the minimum of $\mc S_{g}$
is equal to zero) 4) $\mc S_{g}$ is maximal for a fully mixed state.
5) $\mc S_{g}$ increases under any dephasing operation.

The validity regime for the stochastic laws described in Sec. \ref{subsec: Regime-of-validity},
holds also for the QGCI. Unlike the GCI studied earlier,\textit{ the
adiabats (pure work stage) }in the QGCI\textit{ can include any unitary}
and not only ones in which the energy levels are modified and the
probabilities remain the same. Another important difference with respect
to the stochastic GCI arises from the following property of the Bregman
matrix divergence. Let $\Lambda$ be a diagonal matrix, and $\rho_{\Lambda}$
be the diagonal part of density matrix $\rho$. The Bregman matrix
divergence satisfies
\begin{align}
D_{g}^{B}(\rho,\Lambda) & =D_{g}^{B}(\rho_{\Lambda},\Lambda)+C_{g}(\rho),\label{eq: DB+C}
\end{align}
where $D_{f}^{B}(\rho_{\Lambda},\Lambda)$ is the vector Bregman divergence
(\ref{eq: Bregman def}) of the populations in the $\Lambda$ basis,
and the Bregman coherence measure is 
\begin{equation}
C_{g}(\rho)=\mc S_{g}(\rho_{\Lambda})-\mc S_{g}(\rho)\equiv D_{g}^{B}(\rho,\rho_{\Lambda}).\label{eq: Cbreg}
\end{equation}
This coherence measure has the expected basic properties of a coherence
measure: 1) It is zero for diagonal states 2) It is maximal for the
maximal coherence state $\frac{1}{\sqrt{N}}\sum_{l=1}^{N}\ket l$.
3) It decreases under any dephasing operation. These features follow
from the Schur concavity of $\mc S_{g}(\rho)$. 

Before discussing a certain thermodynamic meaning of $C_{g}$ we point
out that the contractivity condition for isochores now reads
\begin{align}
D_{g}^{B}(\rho_{i},\rho_{0})-D_{g}^{B}(\rho_{f},\rho_{0}) & =D_{g}^{B}(\vec{p}_{i},\vec{p}_{0})-D_{g}^{B}(\vec{p}_{f},\vec{p}_{0})\nonumber \\
 & +C_{g}(\rho_{i})-C_{g}(\rho_{f}).\label{eq: Q contract}
\end{align}
The implication is that even in cases where the stochastic law may
not hold (i.e. the first two terms in (\ref{eq: Q contract}) amount
to a negative number), a significant enough coherence erasure $C_{g}(\rho_{i})-C_{g}(\rho_{f})\ge0$
can be sufficient for making (\ref{eq: Q contract}) positive. Thus
the quantum GCI's can be valid where the stochastic GCI's are not.

\subsection{The thermodynamic operational meaning of various coherence measures}

In the seminal work \cite{PlenioCoherence} coherence measures were
studied from a theoretical point of view without a direct operational
meaning. One of these measures is the quantum relative entropy $D(\rho,\rho_{\Lambda})$
(obtained from (\ref{eq: QBreg}) by choosing $g=-\ln(x)$). In \cite{Anders2015MeasurementWork}
it was shown that $D(\rho,\rho_{\Lambda})$ has a clear quantum thermodynamic
meaning: $TD(\rho,\rho_{\Lambda})$ is the maximal amount of work
that can be extracted from a bath at temperature $T$ using coherence
in the energy basis. $\rho$ is the initial state of system, and $\rho_{\Lambda}$
is the final state of the system. In addition, the initial and final
Hamiltonian are the same. Since the energy of the system does not
change from end to end, the work transferred to the work repository
comes from the bath $-W_{max}=Q_{max}=TD(\rho,\rho_{\Lambda})$. Yet,
to accommodate the second law, this heat-to-work conversion must come
from erasing the coherence. The amount of coherence to be erased is
given by the coherence measure $D(\rho,\rho_{\Lambda})$.

In \cite{Anders2015MeasurementWork} an explicit reversible protocol
was described to achieve this maximal work extraction limit. The protocol
in \cite{Anders2015MeasurementWork} is identical to that described
in appendix II with an additional step where transient pulse (work
stroke, a unitary) is applied to bring $\rho$ to a diagonal form
in the energy basis. Since the only system-bath interaction in the
protocol is an isotherm the result $Q_{max}=TD(\rho,\rho_{\Lambda})$
follows directly form the reversible limit of the second law. Repeating
the same protocol but with the $\alpha$CI we get:

\begin{equation}
\mc Q_{\alpha}^{max}=T^{\alpha}D_{\alpha}^{B}(\rho,\rho_{\Lambda}).\label{eq: Qa max coh}
\end{equation}
Any \textit{irreversible} protocol that generates $\rho\to\rho_{\Lambda}$
will involve less $\mc Q_{\alpha}$ flow to the system. A concrete
irreversible protocol is given in appendix VI. 

In cases where $\mc Q_{\alpha}$ can be related to the standard heat
$\mc Q_{1}$ (e.g. in two-level systems or when condition (\ref{eq: refined high T cond})
holds) the Bregman coherence measure $D_{\alpha}^{B}(\rho,\rho_{\Lambda})$
can produce tighter bounds on the extractable heat and work. For a
two-level system and the irreversible protocol described in Appendix
VI we have numerically verified that a tighter bound on heat extraction
using coherence is obtained compared to the standard second law \cite{Anders2015MeasurementWork}. 

In Appendix VI, we give an example where coherences are used to extract
$Q_{2}$ heat from a thermal bath while exchanging zero standard ($\alpha=1$)
heat. This example illustrates the following point. In reversible
protocols for any $\alpha$, the $Q_{\alpha}$ are completely determined
by the initial state and the temperatures of the bath. In \textit{irreversible}
protocols it is possible to extract different portions of the maximal
$Q_{\alpha}$ determined by the GCI. Thus irreversible protocols offer
more flexibility in manipulating energy distribution via thermal interaction. 

In future studies it is interesting to look for additional quantum
features associate with higher moments of the energy. 

\section{Examples and implications\label{sec: Examples}}

Next, we wish to demonstrate by explicit examples that the GCI's provide
useful information and additional constraints on top of that provided
by the standard CI. Finding the highest impact examples is a matter
of long term research. Instead, our goal is to show that \textit{even
in simple scenarios} the GCI's provide important new input. 

\subsection{High temperature limit }

\begin{figure}
\includegraphics[width=8.6cm]{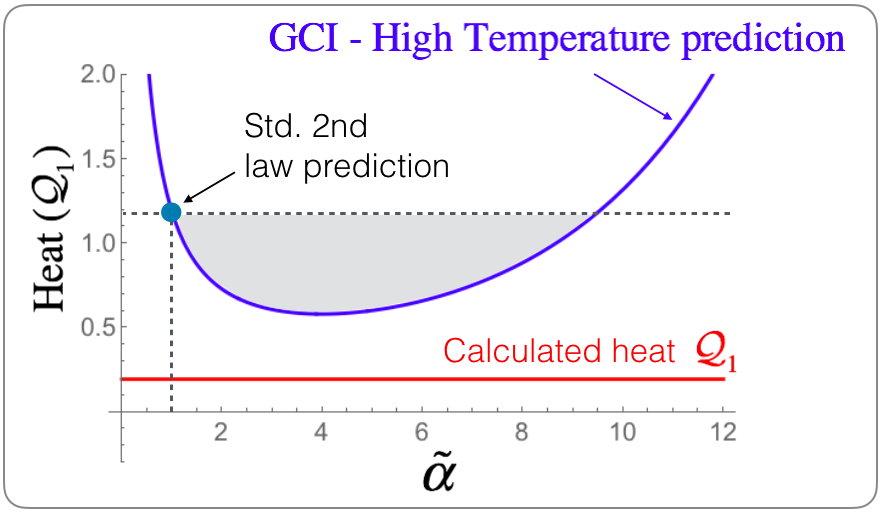}

\caption{\label{fig: High T}High temperature limit of the $\tilde{\alpha}$
impurity-based GCI, applied to a three-level thermalization process
(isochore). For values of $\tilde{\alpha}$ that correspond to the
shaded area, the GCI (\ref{eq: tsalis high T isochore}) imposes a
tighter constraint on the standard $\protect\mc Q_{1}$ heat compared
to the standard second law $\tilde{\alpha}=1$ (dashed-horizontal
line).}
\end{figure}

To illustrate the advantage of (\ref{eq: tsalis high T isochore})
over the standard CI we consider an isochore that eventually fully
thermalizes the system. The chosen energy levels are $E=\{-\frac{1}{2},0,\frac{1}{2}\}$,
the initial state is $p_{0}=\{\frac{1}{2},\frac{1}{2},0\}$, and the
temperature is $T=3$. Figure \ref{fig: High T} shows that there
is a regime of $\tilde{\alpha}$ where the $\tilde{\alpha}$CI provide
a tighter bound on the heat. Of course,  (\ref{eq: tsalis high T isochore})
can be used only when condition (\ref{eq: refined high T cond}) holds.
As discussed in Sec. \ref{subsec: High-temperature-limit} this example
motivates the study of new information measures for bounding the heat
transfer in irreversible processes. It also illustrates the point
that even though the GCI's typically provide predictions on new quantities
such as $\mc Q_{\alpha}$, in some scenarios we can use the GCI's
to learn on standard quantities such as the standard heat. The GCI's
predictions on the heat can be better than the prediction of the standard
CI.

\subsection{Zero standard heat thermodynamics}

\subsubsection{Irreversible case}

Let us recall the zero heat scenario described in the introduction,
and see what insights the $\alpha$CI can provide. We consider the
irreversible case of a simple isochore. Our system has three or more
non-degenerate levels. When connected for a long time to a bath with
inverse temperature $\beta$ (isochore) the final energy of the system
is $\left\langle H\right\rangle _{\beta}=\sum_{j}p_{\beta,j}E_{j}$
regardless of the initial condition ($\vec{p}_{\beta}$ is the thermal
Gibbs state). Now we choose an initial condition $\vec{p}_{i}\neq\vec{p}_{\beta}$
that satisfies $\left\langle H\right\rangle _{i}=\sum_{j}p_{i,j}E_{j}=\left\langle H\right\rangle _{\beta}$,
and we let the system reach $\vec{p}_{\beta}$ (or close enough to
it for all practical purposes). As a result $\left\langle H\right\rangle _{i}=\left\langle H\right\rangle _{f}=\left\langle H\right\rangle _{\beta}$.
Since it is an isochore, there is no work in this scenario, and therefore
$Q=\Delta\left\langle H\right\rangle -W=\Delta\left\langle H\right\rangle =0$.
In this case the standard second law (CI) yields
\begin{equation}
\Delta S\ge0.
\end{equation}
This, however, is a trivial and mathematical statement that can be
obtained even without the CI. The thermal state has the maximal amount
of entropy for a fixed average energy. Since the input and output
state have the same energy, and the output state is thermal, it follows
that the entropy of the input state must be lower, and we get $\Delta S\ge0$.
Moreover, one of the key ideas in thermodynamics is the connection
between entropy (information) and energy, and the second law provides
no such connection in this case. 

Now, let us apply the $\alpha$CI. The second order heat is in general
non-zero. Even if it is zero in some specific case, there is a higher
order heat that is different from zero. For $Q=0$ isochores 
\begin{equation}
\mc Q_{2}=\Delta\left\langle (H-F)^{2}\right\rangle =\Delta\left\langle (H-\left\langle H\right\rangle )^{2}\right\rangle .
\end{equation}
Thus, in $Q=0$ isochores, $\mc Q_{2}$ is the change in the variance
of the energy distribution. In contrast to the $\Delta S\ge0$ for
$\alpha=1$ (CI), the $\alpha=2$ GCI, gives us information on the
energy variance change
\begin{equation}
\Delta\left\langle (H-\left\langle H\right\rangle )^{2}\right\rangle \le T^{2}\Delta\mc S_{2}.\label{eq: irrev examp}
\end{equation}
Full thermalization was used for clarity. Equation (\ref{eq: irrev examp})
is equally valid in cases where the initial state undergoes partial
thermalization that satisfies $Q=0$. From this example we see that
the standard thermodynamic quantities such as the entropy and the
average energy are not sufficient for the thermodynamic description
of zero heat processes, and other quantities such as $\mc H_{\alpha}$
and $\mc S_{\alpha}$ are needed.

It is interesting if in second order phase transitions where the latent
heat is zero, there is a non zero $\alpha$ latent heat which is related
to changes in $\mc S_{\alpha}$.

\subsubsection{Reversible case}

Consider the thermodynamic state preparation scenario where the goal
is to transform a state $p_{i}$ to some other state $p_{f}$ (see
Sec. \ref{subsec: Single-bath-forms and state prep} and Appendix
II). If the protocol is reversible then according to the standard
second law the heat cost is $Q=\mc Q_{1}=T[S(\vec{p}_{f})-S(\vec{p}_{i})]$
where $S=\mc S_{1}$ is the Shannon entropy. 

In systems with three levels or more, there are different distributions
that have the same entropy. In particular, it is possible to have
$S(\vec{p}_{f})=S(\vec{p}_{i})$ where $\vec{p}_{f}\neq\vec{p}_{i}$
and $\vec{p}_{f}$ is not a permutation of the initial $\vec{p}_{i}$.
Figure \ref{fig: S curves} shows the $S$ and $\mc S_{2}$ curves
of the state $p=(1-x/2-x/4,x/2,x/4)$ as a function of the parameter
$x$. The lower horizontal line connects two states that have the
same $S$, even though they are not related by permutation (physically,
permutation is an adiabat). 

As an example, we look at reversible state preparation $\vec{p}_{i}\to\vec{p}_{f}$
where $\vec{p}_{i}$ and $\vec{p}_{f}$ are taken from Fig. \ref{fig: S curves}
and they satisfy $\mc S_{1}(\vec{p}_{f})=\mc S_{1}(\vec{p}_{i})$.
Since it is not a simple permutation, then a bath must be involved
in order to change the values of the probabilities. The upper curve
clearly shows that $\mc S_{2}(\vec{p}_{f})\neq\mc S_{2}(\vec{p}_{i})$.
Although in the interaction with bath $\mc Q_{1}=0$, the second order
heat is non zero $\mc Q_{2}=T^{2}[\mc S_{2}(\vec{p}_{B})-\mc S_{2}(\vec{p}_{A})]\neq0$.
From this example, it is now clear that the bath pays an energetic
price in order to modify the population distribution. That alone is
not a surprising statement, but the $\alpha$CI's \textit{quantify
}this energetic price and\textit{ relate it to information measures}
in the spirit of the standard second law.
\begin{center}
\begin{figure}
\includegraphics[width=8.6cm]{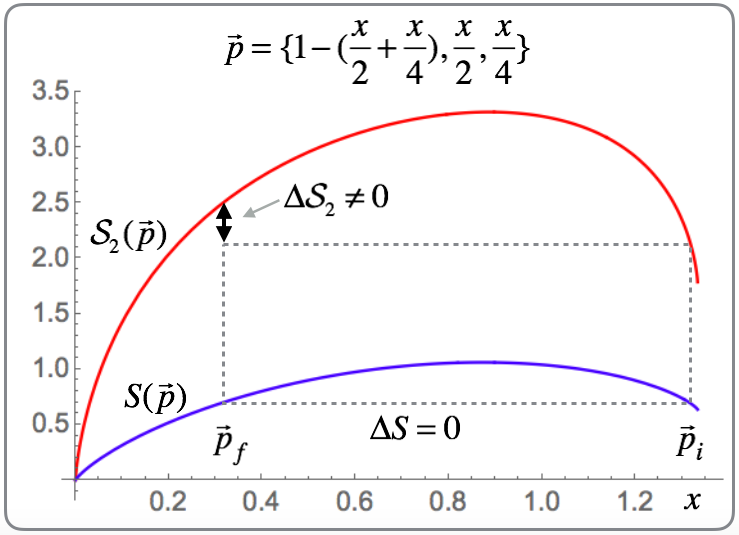}

\caption{\label{fig: S curves}In systems with three or more levels it is possible
to have very different distributions with the same Shannon entropy
$S$. Thus, it is possible to have transformations for which the Shannon
entropy does not change $S(\vec{p}_{f})-S(\vec{p}_{i})=0$, while
other information measures such as $\protect\mc S_{2}$ shown in the
figure, do change $\protect\mc S_{2}(\vec{p}_{f})-\protect\mc S_{2}(\vec{p}_{i})\protect\neq0$. }
\end{figure}
\par\end{center}

\subsection{Otto engine example: tighter than the second law\label{subsec: Tighter-than-2nd}}

The last two examples have explicitly used the energy-information
form the GCI. Next, we want to show that the GCI can also lead to
strong results when using the periodic information-free form of the
GCI (\ref{eq: periodic form}). 

It is not always sensible to compare the predictions of CI's with
different $\alpha$ since generally they contain information on different
observables of the system. However, there are some interesting exceptions.
One is the high temperature limit studied in Sec. \ref{subsec: High-temperature-limit}.
Another exception occurs in two-level systems where all orders of
heat are related to each other. Hence, any $\alpha$CI can be used
to make a prediction on the standard $\mc Q_{1}$ heat. 

\begin{figure}
\includegraphics[width=8.6cm]{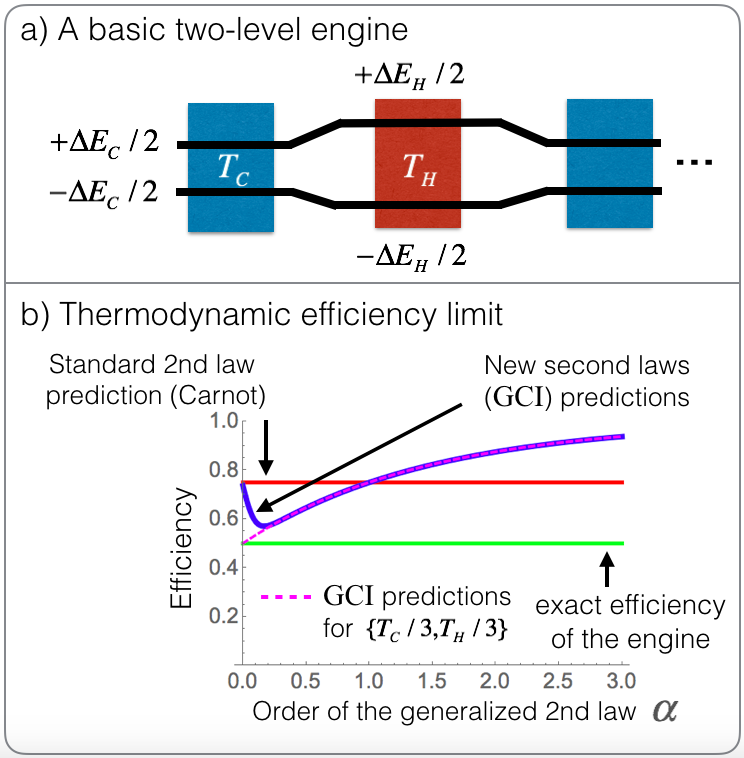}

\caption{\label{fig: Otto TLS}For a two-level four-stroke \textit{irreversible}
engine (a), the $\alpha<1$ order Clausius inequalities provides a
tighter bound on the efficiency (blue curve in (b)), compared to the
$\alpha=1$ standard Carnot efficiency bound (red line). The green
line shows the exact efficiency of this engine. Remarkably, when both
temperatures are decreased by the same factor (three in this example),
the new $\alpha$CI bound (magenta) converges to the exact efficiency
even though the machine is irreversible.}
\end{figure}

From the $\alpha$ order heat definition (\ref{eq: Qa def}) it follows
that for isochore in a two-level system with energies $\{E_{1},E_{2}\}$:
\begin{equation}
\mc Q_{1}=\mc Q_{\alpha}\frac{E_{2}-E_{1}}{(E_{2}-F)^{\alpha}-(E_{1}-F)^{\alpha}}.\label{eq: QaQ1TLS}
\end{equation}
Using it the $\alpha$CI we get that
\begin{equation}
\mc Q_{1}\le T^{\alpha}\Delta\mc S_{\alpha}\frac{E_{2}-E_{1}}{(E_{2}-F)^{\alpha}-(E_{1}-F)^{\alpha}}.\label{eq: Qa to Q1 estimate}
\end{equation}
Now that for two-level system the GCI's give predictions on the standard
heat just like the regular second law we can compare them and see
which one is tighter. As an example we consider the elementary heat
machine shown in Fig. \ref{fig: Otto TLS}a. It is a four-stroke Otto
machine with two levels as a working fluid. In the first stroke the
system is cooled in an isochoric process (levels are fixed in time).
In the second stroke work is invested. The third stroke is a hot isochore
and the fourth is a work extraction stroke. 

The efficiency of this machine is not difficult to calculate. However,
our goal in this example is not to provide simpler methods for evaluating
the efficiency, but to show to what extent thermodynamics puts a restriction
on the efficiency of such an elementary device. The simplicity of
the device shows that the impact of the GCI is not limited to complicated
setups. Moreover in the low temperatures limit $T_{c}\ll E_{c,3}-E_{c,2},T_{h}\ll E_{h,3}-E_{h,2}$
any multi-level Otto machine (without level crossing in the adiabats)
can be accurately modeled as a two-level Otto engine since the third
level population is negligible.

Using (\ref{eq: periodic form}) and (\ref{eq: QaQ1TLS}) we get the
efficiency bound
\begin{equation}
\eta\le\eta_{\alpha CI}=1-\frac{T_{c}^{\alpha}}{T_{h}^{\alpha}}\frac{\frac{E_{c,2}-E_{c,1}}{(E_{c,2}-F_{c})^{\alpha}-(E_{c,1}-F_{c})^{\alpha}}}{\frac{E_{h,2}-E_{h,1}}{(E_{h,2}-F_{h})^{\alpha}-(E_{h,1}-F_{h})^{\alpha}}}
\end{equation}

The parameters are $E_{c,2}-E_{c,1}=1$, $E_{h,2}-E_{h,1}=2$, $T_{c}=0.03$,
$T_{h}=0.12$ and $F_{c(h)}=-T_{c(h)}\ln\sum\exp(-\beta_{c(h)}E_{j}^{c(h)})$
are the standard free energies. In Fig. \ref{fig: Otto TLS}b we see
the actual efficiency (green line) of the engine, the Carnot bound
$1-T_{c}/T_{h}$ from the standard second law (red line), and the
GCI prediction for various $\alpha$ (blue curve). For $\alpha<1$
the $\alpha$CI prediction is significantly tighter compared to the
standard Carnot bound. Since this machine is \textit{irreversible
}it is consistent to have a bound that is tighter than the Carnot
efficiency. 

When operating the same machine with colder temperatures $\{T_{c}',T_{h}'\}=\{T_{c},T_{h}\}/3$
the actual efficiency remains as it was before with $T_{c}$ and $T_{h}$
(Otto engine with uniform compression \cite{RUswap}). The Carnot
bound $\eta_{Carnot}=1-T_{c}'/T_{h}'=1-T_{c}/T_{h}$ also remains
the same. Yet, as shown by the dashed-magenta line in Fig. \ref{fig: Otto TLS}b,
the $\alpha$CI efficiency bound converges to the actual efficiency
for $\alpha\to0$. It is both surprising and impressive that thermodynamic
laws can predict the exact efficiency of an \textit{irreversible}
device. In appendix VII we show analytically that the $\alpha$CI
(\ref{eq: CI ge 0}) becomes tight in low temperatures
\[
T_{c}'\ll E_{c,2}-E_{c,1},T_{h}'\ll E_{h,2}-E_{h,1}
\]
and $\alpha\ll1$ for Otto engines. This result is quite remarkable:
we get a thermodynamic equality even though the machine is not in
the reversible regime (where the crossover to the refrigerator takes
place $T_{c}'/T_{h}'=\Delta E_{c}/\Delta E_{h}$ \cite{RUswap}) or
in the linear response regime $T_{h}'-T_{c}'\ll T_{c}'$.

Next we wish to show how the $\alpha$CI can be used to study a new
type of heat machine in which the standard second law does not provide
any useful information. 

\subsection{Zero heat and half-zero heat machines}

The flow of higher order heat in heat machines is a fascinating subject
that goes beyond the scope of the present paper. However, to motivate
this research direction, we describe a basic Otto ``refrigerator''
that reduces the energy variance of the cold bath without exchanging
any (averaged) energy with it, i.e. $\mc Q_{1,c}=0$. We find that
the performance of such machines \textit{is not} limited by the standard
$\alpha=1$ second law. In contrast, the $\alpha=2$ $\alpha$CI \textit{does
put} a concrete bound on the performance.

\begin{figure}
\includegraphics[width=8.6cm]{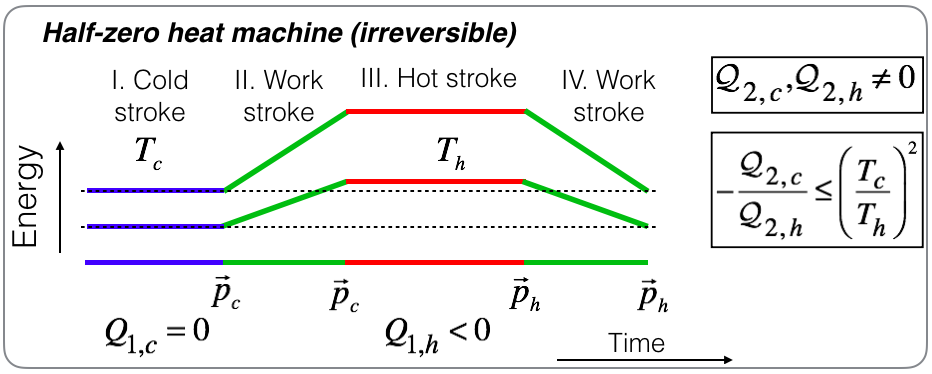}\caption{\label{fig: Zero heat machine} This ``half zero'' heat machine
reduces the energy variance of the cold bath ($-\protect\mc Q_{2,c}\le0$)
without changing the average energy of the bath $\protect\mc Q_{1,c}=0$.
The standard Clausius inequality ($\alpha=1$) only predicts that
the energy goes into the hot bath $\protect\mc Q_{1,h}\le0$ but gives
no information on the changes in the cold bath (which concerns the
main functionality of the device). In contrast, the $\alpha=2$ GCI
(\ref{eq: CI ge 0}) predicts $-\protect\mc Q_{2,c}/\protect\mc Q_{2,h}\le\left(T_{c}/T_{h}\right)^{2}$.
In this example $-\protect\mc Q_{2,c}/\protect\mc Q_{2,h}\simeq0.234$
while the GCI prediction is $\left(T_{c}/T_{h}\right)^{2}=0.25$. }
\end{figure}

The machine we use is a four-stroke Otto machine (see Fig. \ref{fig: Zero heat machine}a)
that interacts with a cold bath in stroke I, and with a hot bath in
stroke III. Some external work is applied to generate the adiabats
in Stroke II and IV. The temperatures and the choice of energy levels
needed to achieve $\mc Q_{1,c}=0$ are given in Appendix VIII.

The variance reduction in the cold bath is based on the fact that
during the cold stroke $\mc Q_{2,c}>0$. Under the conditions in Appendix
III, this implies that the bath experiences an energy variance reduction
of $-\mc Q_{2,c}$. As pointed out earlier, in isochores with $\mc Q_{1}=0$,
$\mc Q_{2}$ is equal to the change in energy variance. 

While the energy flow (first order heat) to the cold bath is zero,
the flow to the hot bath is not zero. From the standard $\alpha=1$
CI for periodic operation we get
\begin{equation}
-Q_{h}/T_{h}\ge0,
\end{equation}
which means that heat enters the hot bath as expected. Since the heat
is zero for only one of the two baths, we call this device a ``half-zero
heat machine''. The energy that flows to the hot bath comes only
from the work. This $\alpha=1$ result is plausible, but it provides
no information on the changes in the cold bath, which concerns the
main functionality of the device. On the other hand, the $\alpha=2$
CI (or other $\alpha$CI) for periodic operation yields
\begin{equation}
-\frac{\mc Q_{2,c}}{\mc Q_{2,h}}\le\left(\frac{T_{c}}{T_{h}}\right)^{2},\label{eq: 2CI machine}
\end{equation}
in cases where $\mc Q_{2,h}\le0$. If $\mc Q_{2,h}$$>0$, the inequality
sign has to be reversed. For the numerical values described in Appendix
VIII, we get $-\mc Q_{2,c}/\mc Q_{2,h}\cong0.234$, while the $\alpha=2$
CI sets a bound of $\left(\frac{T_{c}}{T_{h}}\right)^{2}=0.25$. By
taking lower temperatures it is easy to approach the equality in (\ref{eq: 2CI machine}),
but then $\mc Q_{2,c}$ becomes very small. We conclude that the $\alpha=2$
GCI puts a realistic restriction on the performance of this machine.
It determines what is the minimal amount of $\mc Q_{2}$ the hot bath
must gain to remove $Q_{2,c}$ from the cold bath. This examples shows
that there are cases where the GCI's can provide information which
is more important than that of the standard second law. 

By using reversible state preparation protocol (\ref{eq: WR berg}),
it is possible to construct a ``full-zero heat machines'' where
both $\mc Q_{1,c}$ and $\mc Q_{1,h}$ are equal to zero (see Appendix
VIII). In such machines (\ref{eq: 2CI machine}) becomes an equality. 

The utility and practical value of such machines are outside the scope
of the present paper. Here, the goal is only to show that with the
$\alpha$CI, thermodynamics still imposes limitations on performance
in such scenarios.

\section{Concluding remarks\label{sec:Concluding-remarks}}

This paper presents generalized Clausius inequalities that establish
new connections between information measures, energy moments, and
the temperatures of the baths. As demonstrated, these laws lead to
concrete new predictions in various physical setups. The energy-information
structure of the GCI's separates it from other extensions of the second
law. In particular the GCI's become equalities in reversible processes.

Since the GCI's deal with non-extensive variables (higher moments
of the energy) the information measures associated with these observables
are not extensive as well. The scaling of the GCI with the system
size in different setups is a fascinating topic that warrants further
study. There are two possibilities: 1) All the GCI become trivial
(predict $0\ge0$) in the macroscopic limit. This would imply that
the GCI's are unique to the microscopic domain. 2) In the second scenario
the GCI would provide information on tiny changes in macroscopic systems.
Both alternatives will extend our understanding of thermodynamics
and its scope.

The GCI's have revealed some unexpected features both in hot temperatures,
and in cold temperatures. The hot limit provided better information
measures for estimation of standard heat. In cold temperatures we
have seen that the GCI's can predict the exact efficiency of an engine
even though the engine is irreversible. These interesting GCI's features
should be further explored. 

The regime of validity of the new laws can be smaller than that of
the regular second law (which also has a regime of validity e.g. lack
of initial system-bath correlation). It is a reasonable trade-off:
the validity regime is potentially smaller but more information and
more thermodynamic restrictions are available. In this work only part
of the GCI validity regime has been mapped. The validity regime can
be formulated in terms of ``allowed operations''. We believe it
is highly important to understand and map the full regime of validity.
Based on numerical checks we conjecture that the regime of validity
is significantly larger than the one we were able to deduce analytically
at this point. Moreover, it is possible that some adaptations and
refinements of the framework presented here will lead to a larger
regime of validity.

Within this known regime of validity we provided explicit examples
where the generalized Clausius inequality gives tighter and more useful
constraints on the dynamics compared to the standard second law. A
quantum extension was presented and used for providing a thermodynamic
interpretation to various coherence measures. All the above-mentioned
findings justify further work on this topic. The main goals are: 1)
Finding additional predictions 2) Explore various limits (large $\alpha$,
macroscopic limit, cold temperatures, etc.) 3) Extending the mapped
regime of validity. 

Moreover, it is interesting to extend the GCI formalism to other physical
scenarios. For example, include chemical potentials in the GCI, and
apply it to thermoelectric devices and molecular machines \cite{Leigh2015RiseMachines}.
Another interesting option is to extend our findings to continuous
distributions, and study dynamics of classical particles in a box
(gas) from the point of view of the GCI's. It is also interesting
to study latent $\alpha$ heat in various phase transitions. 
\begin{acknowledgments}
The author is indebted to Prof. Christopher Jarzynski for stimulating
discussions and useful suggestions. Part of this work was supported
by the COST Action MP1209 'Thermodynamics in the quantum regime'. 
\end{acknowledgments}

\section*{Appendix I - From Bregman divergences to generalized Clausius equalities}

\begin{figure}
\includegraphics[width=8.6cm]{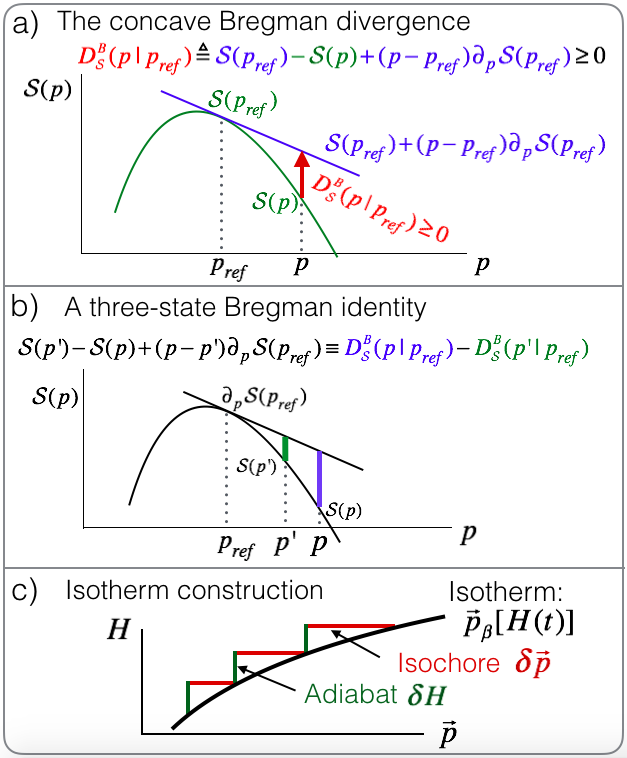}

\caption{\label{fig: Bregman}(a) The (concave) Bregman divergence has a simple
geometric interpretation. It is the difference between the linear
extrapolation of a concave function (blue) and the actual function
(green). (b) A the three-state Bregman identity and its geometric
interpretation. The right hand side determines the regime of validity
of the generalized Clausius inequalities. Geometrically, it is the
length of the blue line (initial divergence with respect to the reference)
minus the length of the green line (final divergence with respect
to the reference). If the final state is closer to the reference state
(but does not overshoot to the other side) then the divergence difference
is guaranteed to be positive as needed for the generalized Clausius
inequalities. (c) By using concatenation of isochores and adiabats
it is possible to construct various thermodynamic protocols that involve
both heat and work such as isotherms. See text for the analysis.}
\end{figure}

The Bregman divergence for a single-variable concave and differentiable
function $\mc S(p)$ in the regime $p\in[0,1]$ is given by \cite{Bregman1967}
\begin{align}
D_{\mc S}^{B}(p,p_{ref}) & =\mc S(p_{ref})-\mc S(p)+(p-p_{ref})(\partial_{p}\mc S)|_{p=p_{ref}},\label{eq: Berg pref 2}\\
D_{\mc S}^{B}(p,p_{ref}) & \ge0,
\end{align}
where $p_{ref}$ is any point in $[0,1]$. The Bregman divergence,
as shown in Fig. \ref{fig: Bregman}a, is the difference between a
concave function and its linear extrapolation from point $p_{ref}$.
Writing (\ref{eq: Berg pref 2}) again with $p\to p'$ and subtracting
(\ref{eq: Berg pref 2}) from it we get
\begin{align}
\mc S(p')-\mc S(p)-(p'-p)\partial_{p}\mc S(p_{ref}) & =D_{\mc S}^{B}(p,p_{ref})\nonumber \\
 & -D_{\mc S}^{B}(p',p_{ref}).\label{eq: gen berg diff}
\end{align}
This equation also has a geometrical interpretation as shown in Fig.
\ref{fig: Bregman}b. In particular, Fig \ref{fig: Bregman}b shows
that when the final state is closer to the reference state from the
same side, then the RHS of (\ref{eq: gen berg diff}) is positive.
To apply this for a probability distribution $\{p_{j}\}_{j=1}^{N}$
of $N$-level system we define
\begin{equation}
\mc S(\vec{p})=\sum_{j=1}^{N}\mc S(p_{j}),
\end{equation}
and obtain a vector generalization of (\ref{eq: gen berg diff})
\begin{align}
\mc S(\vec{p}\:')-\mc S(\vec{p})-(\vec{p}\:'-\vec{p})\cdot\nabla\mc S(\vec{p}_{ref}) & =D_{\mc S}^{B}(\vec{p},\vec{p}_{ref})\nonumber \\
 & -D_{\mc S}^{B}(\vec{p}\:',\vec{p}_{ref}).
\end{align}
To apply this to thermodynamics we set $\vec{p}\:'$ to be the final
state of the system and $\vec{p}$ to be its initial state. The term
$(\vec{p}\:'-\vec{p})\cdot\nabla\mc S(\vec{p}_{ref})$ is a difference
of two terms of the form $\sum p_{j}(\nabla\mc S(\vec{p}_{ref}))_{j}\triangleq\left\langle \nabla\mc S(\vec{p}_{ref})\right\rangle $
so $(\vec{p}\:'-\vec{p})\cdot\nabla\mc S(\vec{p}_{ref})$ describes
the change in the expectation value of the operator $\nabla\mc S(\vec{p}_{ref})$.
Note that the operator does not depend on the initial and final distributions
but only on the reference distribution $\vec{p}_{ref}$ that we will
choose shortly. With this notations
\begin{equation}
\Delta\mc S-\Delta\left\langle \nabla\mc S(\vec{p}_{ref})\right\rangle =D_{\mc S}^{B}(\vec{p},\vec{p}_{ref})-D_{\mc S}^{B}(\vec{p}\:',\vec{p}_{ref}).\label{eq: aCI pref}
\end{equation}
This is still an identity that has nothing to do with thermodynamics.
To get a Clausius-like inequality we want the RHS of (\ref{eq: aCI pref})
to be positive for thermodynamic process such as isochores, adiabats,
and isotherms. As discussed in the main text, isochores can be used
as a starting point. If the bath has a single fixed point so that
the map it induces on the system $\vec{p}_{f}=M(\vec{p}_{i})$ satisfies
$M(\vec{p}_{f.p.})=\vec{p}_{f.p.}$, we choose $\vec{p}_{ref}=\vec{p}_{f.p.}$.
Since the goal of the bath is to bring the system closer to the fixed
point, it is plausible that the RHS will be positive. However, although
the bath may bring the state closer to the fixed point by some divergence
measures, such as the relative entropy, it is not guaranteed that
it will bring it closer when using the $D_{\mc S}^{B}$ as a proximity
measure. Thus, one has to explore the regime of validity and check
whether the given thermalization mechanism is ``contractive under
$D_{\mc S}^{B}$'' (the RHS of (\ref{eq: aCI pref}) is positive).
This is done in Sec. \ref{subsec: Regime-of-validity}. 

For a single bath that is connected to all the levels of the system
the fixed point is the thermal state so we set the reference state
to be $p_{ref,j}=p_{\beta,j}=e^{-\beta(E_{j}-F)}$. For the choice
$\partial_{p_{j}}\mc S(\vec{p})=(-\ln p_{j})^{\alpha}$ we get (\ref{eq: Bregman isochore}). 

In addition, our formalism is also applicable to cases where different
baths are connected to different parts of the system. These ``parts'',
that we call manifolds \cite{EquivPRX}, can either be in tensor product
form when the system is composed of several particles, or in a direct
sum form when different levels of the same particle are connected
to baths with different temperatures. Such a scenario is common in
microscopic heat machines (see \cite{EquivPRX,k102,scovil59}). 

The $k$ manifold of the system is associated with the part of the
Hamiltonian $H_{k}=\sum_{j\in\{k\}}E_{j}\ketbra jj$ and it interacts
with a bath of temperature $\beta_{k}$. The fixed point is $p_{ref,j}=p_{\vec{\beta},j}=e^{-\sum_{k}\beta_{k}(H_{k}-F_{k})}$
where $F_{k}$ are chosen so that each manifold has the correct total
probability. If the manifolds share just one state then the fixed
point is unique \cite{EquivPRX}. For the $\alpha$CI choice $\partial_{p}\mc S=(-\ln p)^{\alpha}$
we get

\begin{eqnarray}
\Delta\mc S_{\alpha}-\sum_{k}\beta_{k}^{\alpha}\mc Q_{k,\alpha} & = & D_{\mc S}^{B}(\vec{p},\vec{p}_{\vec{\beta}})-D_{\mc S}^{B}(\vec{p}\:',\vec{p}_{\vec{\beta}}),\label{eq: gen CID}\\
\mc Q_{k,\alpha} & = & \Delta\left\langle (H-F_{k})^{\alpha}\right\rangle .\label{eq: Qk alpha}
\end{eqnarray}
Equation (\ref{eq: gen CID}) refers to isochores. For more general
processes that include adiabats, isochores, and isotherms the derivation
presented in the main text has to be repeated. In the regime of validity
discussed in Sec. \ref{subsec: Regime-of-validity}, equation (\ref{eq: gen CID})
yields
\begin{equation}
\Delta\mc S_{\alpha}-\sum_{k}\beta_{k}^{\alpha}\mc Q_{k,\alpha}\ge0.\label{eq: gen CI}
\end{equation}

Finally, we want to show how that the equality in the $\alpha$CI
$\Delta\mc S_{\alpha}=\beta^{\alpha}\mc Q_{\alpha}$ is obtained for
isotherms by using a concatenation of isochores and adiabats. This
is an alternative derivation to (\ref{eq: Q iso proof}). See \cite{anders2013thermodynamics}
for a similar analysis. Yet, here we carry out the calculation for
the GCI and not only for the standard CI. Figure 5c shows a concatenation
that approximates an isotherm $p_{\beta}[H(t)]$ (black curve). In
the limit where the step size $\delta\vec{p}$ goes to zero the concatenation
converges to the isotherm. We start at equilibrium and then perform
a small change in the Hamiltonian $\delta H$ (green line). $\vec{p}$
remained fixed in this process so this is an adiabat (in particular,
the entropy has not changed, and not heat was exchanged with the bath).
So for the adiabat we get $\delta\mc S_{\alpha}=0,\delta\mc Q_{\alpha}=0$.
Next we perform an isochore (red line) all the way to the ideal isotherm
line. From (\ref{eq: Bregman isochore}) we get that for full thermalization
($\vec{p}_{f}=\vec{p}_{\beta}$) isochores 
\begin{equation}
\delta\mc S_{\alpha}-\beta^{\alpha}\mc Q_{,\alpha}=D_{\mc S}^{B}(\vec{p}_{\beta}-\delta\vec{p},\vec{p}_{\beta})
\end{equation}
By definition the term $\mc Q_{,\alpha}$ is linear in $\delta\vec{p}$.
For the RHS we use a general property of the Bregman divergence $D_{\mc S}^{B}(\vec{p}-\delta\vec{p},\vec{p})=O(\delta\vec{p}^{2})$.
This holds since $D_{\mc S}^{B}(\vec{p},\vec{p})=0$ and $D_{\mc S}^{B}(\vec{p}-\delta\vec{p},\vec{p})>0$
when $\delta\vec{p}\neq0$. If there was a linear term, then by taking
$\delta\vec{p}\to-\delta\vec{p}$ the divergence would have become
negative when $\delta\vec{p}$ is very small. Figure 5a offers another
way of understanding this property. The Bregman divergence is obtained
from the function $\mc S$ by subtracting its linear extrapolator,
and therefore it no longer has a linear term. Due this property we
find that for the first stair in this staircase
\begin{equation}
\delta\mc S_{\alpha}^{(1)}=\beta^{\alpha}\delta\mc Q_{,\alpha}^{(1)}+O(\delta\vec{p}^{2}).
\end{equation}
Repeating this for the $l$ stair and summing we find
\begin{equation}
\sum_{l}\delta\mc S_{\alpha}^{(l)}=\beta^{\alpha}\delta\mc Q_{,\alpha}^{(l)}+\sum_{l}O(\delta\vec{p}^{2}).
\end{equation}
Since $\delta\mc Q_{,\alpha}^{(l)}=O(\delta\vec{p})$ then $\delta\mc S_{\alpha}^{(l)}$
must also be $O(\delta\vec{p})$ to balance the equation. The term
$O(\delta\vec{p}^{2})$ becomes negligible in the limit $\delta\vec{p}\to0$
and we get $\Delta\mc S_{\alpha}=\beta^{\alpha}\mc Q_{\alpha}$. 

\section*{Appendix II - Reversible state preparation}

In this appendix we derive (\ref{eq: WR berg}). Since we are interested
now in \textit{reversible} state preparation we can choose any reversible
protocol that achieves the transformation $\{\vec{p}_{i},H_{i}\}\to\{\vec{p}_{f},H_{f}\}$.
\begin{figure}
\includegraphics[width=8.6cm]{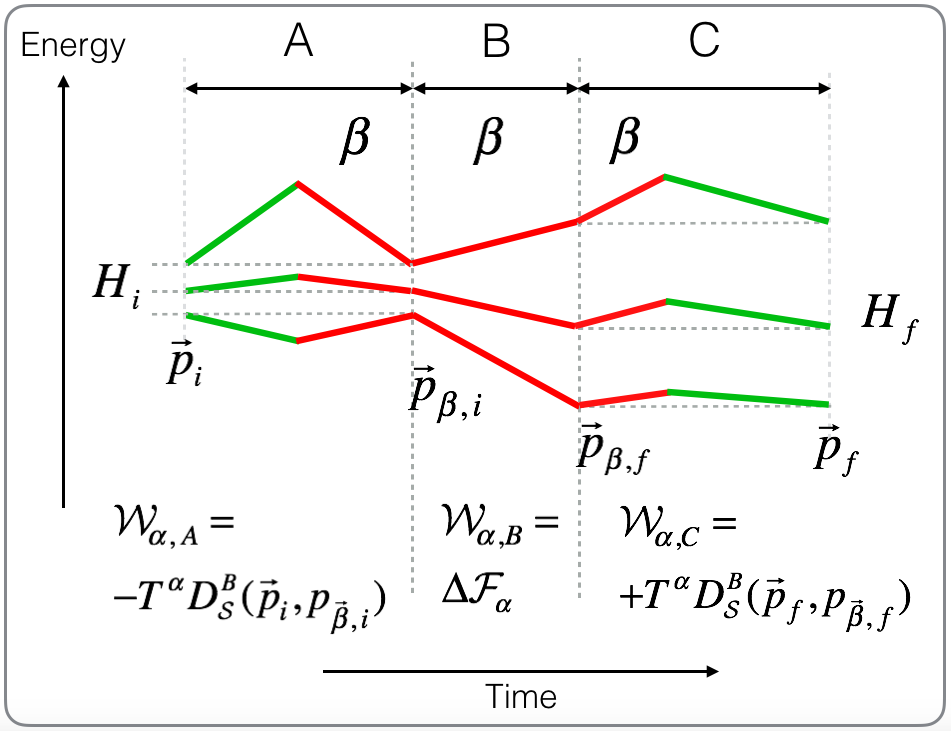}

\caption{\label{fig: rev protocol}A three-stage reversible state-preparation
protocol $\{\vec{p}_{i},H_{i}\}\to\{\vec{p}_{f},H_{f}\}$. Stage A
changes the populations but the end point Hamiltonian is the same
as the initial Hamiltonian. Stage B is an isotherm in which $H_{i}$
is changed to $H_{f}$. Finally, stage C changes the thermal state
to the desired final state. This protocol is identical to state preparation
with $\alpha=1$ but we monitor other quantities in the process ($\protect\mc W_{\alpha}$
and $\protect\mc Q_{\alpha}$). In this protocol the GCI's become
equalities and they give the exact amount of extracted $\protect\mc W_{\alpha}$
and $\protect\mc Q_{\alpha}$. }
\end{figure}
Figure \ref{fig: rev protocol} shows the protocol we chose for the
derivation. Stage A implements the transformation $\{\vec{p}_{i},H_{i}\}\to\{\vec{p}_{\beta,i}=e^{-\beta(H_{i}-F_{i})},H_{i}\}$.
Stage B is an isotherm $\{\vec{p}_{\beta,i},H_{i}\}\to\{\vec{p}_{\beta,f}=e^{-\beta(H_{f}-F_{f})},H_{f}\}$,
and in stage C $\{\vec{p}_{\beta,f},H_{f}\}\to\{\vec{p}_{f},H_{f}\}$
is carried out. Starting with stage A we use (\ref{eq: Berg pref 2})
and (\ref{eq: S def}) to write
\begin{eqnarray}
T^{\alpha}\Delta\mc S_{\alpha,A}-\Delta\mc H_{\alpha,A} & = & T^{\alpha}D_{\mc S}^{B}(\vec{p}_{f},p_{\vec{\beta},f}).
\end{eqnarray}
Using $\mc Q_{\alpha,A}^{R}=T^{\alpha}\Delta\mc S_{\alpha,A}$ and
the first law $\mc Q_{\alpha}^{R}-\Delta\mc H_{\alpha,A}=-\mc W_{\alpha,A}$,
the reversible work extracted in the transformation $\{\vec{p}_{i},H_{i}\}\to\{\vec{p}_{\beta,i},H_{i}\}$
is
\begin{equation}
\mc W_{\alpha,A}=-T^{\alpha}D_{\mc S}^{B}(\vec{p}_{i},p_{\vec{\beta},i}).
\end{equation}
Similarly, in stroke C (just the inverse of protocol used in A) the
reversible work in $p_{\vec{\beta},f}$ to $\vec{p}_{f}$ is
\begin{equation}
\mc W_{\alpha,C}=+T^{\alpha}D_{\mc S}^{B}(\vec{p}_{f},p_{\vec{\beta},f}).
\end{equation}
The last bit we need is the work in stage B. This is a pure isotherm
so $\mc Q_{\alpha,B}=T^{\alpha}\Delta\mc S_{\alpha,B}$ and $\mc W_{\alpha,B}=[\mc H_{\alpha}(p_{\vec{\beta},f})-\mc H_{\alpha}(p_{\vec{\beta},i})]-T^{\alpha}[\mc S_{\alpha}(p_{\vec{\beta},f})-\mc S_{\alpha}(p_{\vec{\beta},f})]$.
Therefore
\begin{eqnarray}
\mc W_{\alpha,B} & = & \mc F_{\alpha}(\vec{p}_{\beta,f})-\mc F_{\alpha}(\vec{p}_{\beta,i})=\Delta\mc F_{\alpha},\\
\mc F_{\alpha} & = & \mc H_{\alpha}(\vec{p}_{\beta})-T^{\alpha}\mc S_{\alpha}(\vec{p}_{\beta}),
\end{eqnarray}
where $\mc F_{\alpha}$ is the \textit{equilibrium} $\alpha$ free
energy (no relation to the ``$\alpha$ free energy'' in thermodynamic
resource theory). Adding the work contribution from all stages, we
obtain that the total reversible work is given by (\ref{eq: WR berg}). 

\section*{Appendix III - $\alpha$ order heat exchange with the bath}

Our definition of heat and work can be considered as axioms. We define
some observables of the system and put some thermodynamic constraints
on how they can change in the spirit of the standard second law. Yet,
in $\alpha=1$ thermodynamics the heat absorbed by the system is taken
from the bath. To be more accurate, this is not always true since
there might be some additional energy (work) needed to couple the
system to the bath. In weak coupling this energy can be ignored but
also in certain strong coupling cases \cite{RUnonMarkovianEquiv}.

Nevertheless, regardless of where the energy goes, energy conservation
implies that any energy change in the system is associated with an
opposite change in the energy of the surroundings. Unfortunately,
there is no general conservation law for $\mbox{\ensuremath{\mc H}}_{\alpha}$,
so what can we learn on the change in the surroundings from the changes
of $\mbox{\ensuremath{\mc H}}_{\alpha}$ in the system? We first focus
on heat and in particular on isochores, and later discuss a specific
yet important scenario of $\alpha$ work extraction.

\subsection{System-bath $\alpha$ heat flow}

\begin{figure}
\includegraphics[width=8.6cm]{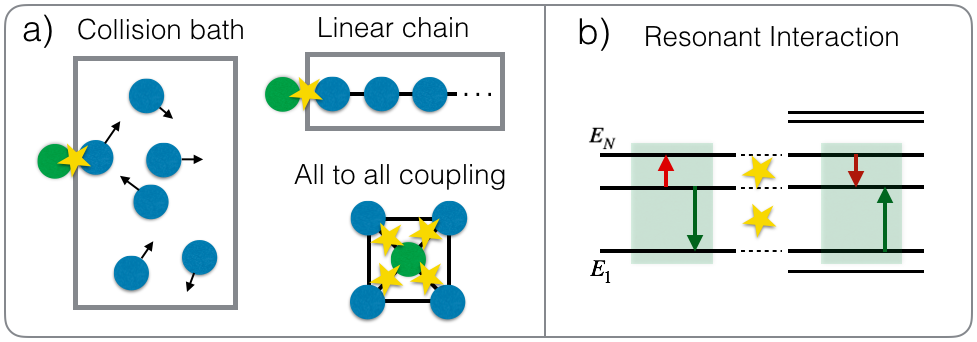}

\caption{\label{fig: baths}(a) Various system-bath configurations, and the
interaction between elements (b). Each energy quanta taken from the
bath particles is given to the system and vice versa. This type of
interaction enables to easily relate changes in energy moments of
the system, to changes in the bath.}
\end{figure}

In our bath setup the bath consists of particles that can interact
with the system and/or with each other (Fig. \ref{fig: baths}a).
The Hamiltonian of the internal degrees of freedom of a particle $k$
in the bath is $H_{b,k}$. We make the following assumption 
\begin{equation}
H_{b,k}=H_{s}+\sum_{m}e_{m}\ketbra mm.\label{eq: Hb Hs-1}
\end{equation}
That is, the bath particle has the same energy levels of the system
plus (or minus) possibly additional levels $\ket m$ (see Fig. \ref{fig: baths}b).
This is a reasonable assumption when the system \textit{resonantly}
interacts with the bath or when the bath particles and the system
particle are of the same species. The setup studied here can describe
various models. Several examples are shown in Fig. \ref{fig: baths}a:
a collision bath model, a linear chain with nearest neighbor coupling,
and all to all coupling geometry. For the interaction of the system
with the bath particles, and of the bath particle between themselves,
we assume a resonant ``flip-flop'' interaction term (Fig. \ref{fig: baths}b)
\begin{equation}
H_{int}=\sum_{j,k,l}c_{jk,l}\sigma_{j,l}^{-}\sigma_{k,l}^{+}+h.c.,\label{eq: Hint}
\end{equation}
where $\sigma_{k,l}^{+}=\ketbra{i+\omega_{l}}i$ is the creation operator
of energy gap $\omega_{l}$ in particle $k$, and $\sigma_{k,l}^{-}=(\sigma_{k,l}^{+})^{\dagger}$
is the corresponding annihilation operator. For simplicity, it is
assumed that the gaps are non-degenerate so specifying $\omega_{l}$
specifies the state $\ket i$ as well. This two-particle interaction
conserves the total \textit{bare} energy $\left\langle H_{tot}\right\rangle =\left\langle H_{s}\right\rangle +\sum_{k}\left\langle H_{b,k}\right\rangle $,
and it is very common in ion traps and in superconducting circuits
(after making a justified rotating wave approximation). Furthermore,
from $[H_{tot},H_{int}]=0$ that leads to energy conservation it also
follows that $H_{tot}^{(\alpha)}$ is conserved where
\begin{equation}
\left\langle H_{tot}^{(\alpha)}\right\rangle \doteq\left\langle H_{s}^{\alpha}\right\rangle +\left\langle H_{b}^{(\alpha)}\right\rangle \doteq\left\langle H_{s}^{\alpha}\right\rangle +\sum_{k}\left\langle H_{b,k}^{\alpha}\right\rangle =\text{const}.\label{eq: H(a)tot}
\end{equation}
Note that $H_{b}^{(\alpha)}$ is equal to $H_{b}^{\alpha}\otimes I\otimes..+I\otimes H_{b}^{\alpha}\otimes..+..$
and not to $(H_{b}\otimes I\otimes..+I\otimes H_{b}\otimes..)^{\alpha}$.
Interaction of the form (\ref{eq: Hint}) can only redistribute energy
(or $\left\langle H^{(\alpha)}\right\rangle $) between system and
bath. Hence, no extra work (or $W^{\alpha}$ work) is needed to couple
the system and the bath. 

When the bath and system particles are of the same species $F_{sys}=F_{bath}$
and it follows from (\ref{eq: H(a)tot}) that $\Delta\left\langle \mc H_{b}^{(\alpha)}\right\rangle =-\Delta\left\langle \mc H_{sys}^{\alpha}\right\rangle $
or alternatively stated 
\begin{equation}
\mc Q_{\alpha,b}=-\mc Q_{\alpha,sys}.
\end{equation}
What happens $F_{s}\neq F_{bath}$? We go back to the motivation of
defining of $\mc H_{sys}^{\alpha}$ as $(H-F)_{sys}^{\alpha}$. This
followed from the Bregman divergence definition and from the fact
that the Gibbs state is a fixed point of any thermalization map created
by the bath. For the bath, these considerations are irrelevant. We
can define $\mc H_{k}^{(\alpha)}$ of the k particle in the bath with
any shift of interest. In particular, we can choose
\begin{equation}
\mc H_{b,k}^{\alpha}=(H_{b,k}-F_{sys})^{\alpha}.
\end{equation}
With this choice we get $\Delta\left\langle \mc H_{b}^{(\alpha)}\right\rangle =-\Delta\left\langle \mc H_{sys}^{\alpha}\right\rangle $.
For other observables such as $\left\langle e^{-(\tilde{\alpha}-1)\beta(H-F)}\right\rangle $
associated with the $\alpha$ impurity (see Sec. \ref{sec:Tsalis and Renyi})
the relation is more straightforward. For any choice of $F_{b}$ we
can use $\mc Q_{\tilde{\alpha}}=\left\langle e^{-(\tilde{\alpha}-1)\beta(H-F)}\right\rangle =e^{(\tilde{\alpha}-1)\beta F}\left\langle e^{-(\tilde{\alpha}-1)\beta H}\right\rangle $
and (\ref{eq: H(a)tot}) and get
\begin{equation}
\mc Q_{\tilde{\alpha}}^{bath}=-e^{-(\tilde{\alpha}-1)\beta(F_{sys}-F_{b})}\mc Q_{\tilde{\alpha}}^{sys}.\label{eq: Q2s Q2b Tsal}
\end{equation}
The relation (\ref{eq: Q2s Q2b Tsal}) holds for isochores. If the
levels change in time a time integration over $\delta Q$ with $F_{sys}(t)$
has to be carried out. 

In the bath models above the flip-flop interactions lead to multiple
conservation laws. Consequently if some energy is exchanged with the
bath and it is not in equilibrium anymore. Moreover, due to the conservation
laws the bath will not equilibriate on its own once disconnected from
the system. This is exactly the reason why it is important to keep
track of higher order heat flow. They quantify the degradation of
the bath. By assuming the bath is in a thermal state with unknown
$\beta$ any knowledge of $\left\langle \mc H_{b}^{(\alpha)}\right\rangle $
can be used to get the correct $\beta$. However if the distribution
is non thermal different $\alpha$ will produce different prediction
for $\beta$. The $\beta$ mismatch is an indication for deviation
of the bath from a thermal distribution.

\subsection{$\alpha$ Work}

We now consider the case of applying a transient unitary on the system
by driving it with $H_{s}(t)$. At the end of the pulse the Hamiltonian
returns to its original value. In such scenarios the change in energy
or in $\mc{H_{\alpha}}$ is only due to work ($\mc W_{\alpha}$).
The problem is that in general it is not clear how a change of $\mc W_{\alpha}$
in the system is related to changes in the work repository. To simplify
things we look at systems where specific two levels resonantly interact
with a work repository. In the semi-classical limit the work repository
is a harmonic oscillator in highly excited coherent state. This scenario
is very useful in quantum heat machines \cite{scovil59,k102,EquivPRX}.
The $\mc W_{\alpha}$ is
\begin{align}
\mc W_{\alpha} & =\sum_{j=\{k,l\}}(E_{j}-F)^{\alpha}dp_{j}\nonumber \\
 & =[(E_{l}-F)^{\alpha}-(E_{k}-F)^{\alpha}]dp_{l}.
\end{align}
Since the regular work is $W=\mc W_{1}=(E_{l}-E_{k})dp_{l}$ we get
a simple relation between work and $\alpha$ order work:
\begin{equation}
W=\frac{(E_{l}-E_{k})}{[(E_{l}-F)^{\alpha}-(E_{k}-F)^{\alpha}]}\mc W_{\alpha}.\label{eq: W Walpha}
\end{equation}
In particular, in high temperature where $\left|F\right|\gg\left|E_{l}\right|,\left|E_{k}\right|$
we get: $W=\frac{(-F)^{1-\alpha}}{\alpha}\mc W_{\alpha}$. Surely,
relation (\ref{eq: W Walpha}) does not provide a sufficient understanding
of the operational effect of $\mc W_{\alpha}$ on the work repository,
and further study on this topic is needed. Nevertheless, (\ref{eq: W Walpha})
is already sufficient to relate standard work to higher order heat
flows $\mc Q_{\alpha}$ in certain classes of machines mentioned above.

Finally, we point out that there are thermodynamic scenarios like
in absorption refrigerators where energy flows only in the form of
heat, and there is no work at all. 

\section*{Appendix IV - resource theory monotones}

Like the generalized CI, thermodynamic resource theory (TRT) \cite{Goold2015review,horodecki2013fundamental,BrandaoPnasRT2ndLaw,GourRTreview,SaiJanetReview}
also puts further restrictions on the interaction of a system with
a thermal bath. Yet, as explained next, the similarities to the present
framework seem to end there (note that the $\alpha$ index used in
TRT has a completely different meaning). Both frameworks have their
merits and deficiencies. In our view, both of them provide different
tools for studying thermodynamic transformations at the microscopic
scale. 

Resource theory is the study of possible transformations from one
state to the other, by using ``free states'' and possibly non-free
states that are considered as a resource \cite{Goold2015review,SaiJanetReview,GourRTreview}.
The free states in TRT are the thermal states. TRT is presently limited
to scenarios with a single thermal bath (single temperature). In standard
thermodynamics for a single bath, the CI can be replaced by the non-equilibrium
free energy inequality
\begin{eqnarray}
\Delta\tilde{F} & \le & 0,\\
\tilde{F} & = & F+TD(p,p_{\beta}).\label{eq: Fneq}
\end{eqnarray}
That is, $\tilde{F}$ is a monotone under certain thermodynamic transformations.
Thermodynamic resource theory states that under ``Thermal operations''
\cite{Goold2015review,SaiJanetReview} the free energy is only one
member of a whole monotone family \cite{BrandaoPnasRT2ndLaw}:
\begin{eqnarray}
\tilde{F}_{\breve{\alpha}} & = & F+TD_{\breve{\alpha}}(\vec{p},\vec{p}_{\beta}),\label{eq: Fren}\\
D_{\breve{\alpha}}(\vec{p},\vec{q}) & = & \frac{\text{sign}(\breve{\alpha})}{\breve{\alpha}-1}\ln\sum_{i}p_{j}^{\breve{\alpha}}q_{j}^{1-\breve{\alpha}}.
\end{eqnarray}
Where $\breve{\alpha}$ is a real number and $D_{\breve{\alpha}}(\vec{p},\vec{q})$
is the Rényi \textit{divergence} (not to be confused with the Rényi
entropy in (\ref{eq: Renyi CI}), or with the Bregman divergence $D_{\bar{\alpha}}^{B}$
related to the Rényi entropy). If the initial state has coherences
in the energy basis, then there are additional constraints \cite{LostaglioRudolphCohConstraint}.
What is remarkable about these \textit{thermodynamic monotones} (\ref{eq: Fren})
is that they provide necessary and \textit{sufficient} conditions
for the existence of a thermal operation. That is, if all the monotones
decrease in the transformation of two energy diagonal density matrices
$\rho\to\sigma$, then a thermal operation that generates the transformation
$\rho\to\sigma$ exists.

These thermodynamic monotones are sometimes referred to in the literature
as ``second laws''. The reasons that support this terminology are:
1) They have to decrease under thermal operation 2) The reduction
of (\ref{eq: Fren}) to the standard (non-equilibrium) free energy
(\ref{eq: Fneq}) in the limit $\breve{\alpha}\to1$. However, in
our view the second law is more than a thermodynamic monotone. Consider
the Clausius \textit{equality }for isochores (\ref{eq: Bregman isochore}).
The RHS is the monotone part of the equality, and it is positive in
the regime of validity described in Sec. \ref{subsec: Regime-of-validity}.
In traditional thermodynamics, it is the LHS that gives thermodynamics
its strength. The LHS deals with thermodynamic quantities. In particular
it has an energy-information structure.

In contrast, in TRT $\tilde{F}_{\breve{\alpha}}$ generally involves
non-integer power of probabilities, and therefore cannot be directly
related to observables. Presently, to the best of our knowledge, an
operational\textit{ thermodynamic} meaning to $F_{\breve{\alpha}}$
for $\breve{\alpha}\neq1$ (with the exclusion of $\tilde{F}_{\infty}$
and $\tilde{F}_{0}$ ) is still lacking (there is an informational
state discrimination interpretation). In \cite{FunoWorkFlucTradeOffRenyi}
the $F_{\breve{\alpha}}$ are used to obtain an interesting $\breve{\alpha}$
independent result on the work fluctuation-dissipation trade-off with
a single bath (``information engine'' scenario - not a multiple
bath heat engine scenario). See also \cite{wilming2017ThirdLawRT,Woods2015EffEng}
for different interesting directions of applying TRT.

More importantly, the TRT constraints (\ref{eq: Fneq}) \textit{do
not} have an energy-information structure. The Rényi divergence $D_{\breve{\alpha}}(\vec{p},\vec{p}_{\beta})$
is a measure of \textit{distinguishability of a state }$\vec{p}$\textit{
from the thermal state $\vec{p}_{\beta}$.} However it is not a measure
of the information content in the distribution $\vec{p}$. For example
it is not even invariant to permutations in $\vec{p}$. 

In summary, it seems that there is very little similarity between
the GCI and TRT. Despite the differences, we hope that the TRT framework
and the generalized CI framework can benefit from each other on the
road to a deeper understanding of thermodynamics of small systems.
As an example for exchange of ideas between the two formalisms, it
is interesting to investigate the following question: we are given
the transformation $\vec{p}\to\vec{q}$ and we find that $D_{\alpha}^{B}(\vec{p},\vec{p}_{\beta})\ge D_{\alpha}^{B}(\vec{q},\vec{p}_{\beta})$
for any $\alpha$, i.e. all the $\alpha$CI are satisfied. In analogy
to thermal operation in TRT, does it imply that $\vec{p}\to\vec{q}$
can always be implemented with the allowed set of operations? The
same question can be posed for other families of GCI's such as the
ones studied in Sec. \ref{sec:Tsalis and Renyi}.

\section*{Appendix V - On the physicality of $\protect\mc S_{\alpha}$}

The goal of this section is not to enter the somewhat philosophical
discussion on the physicality of information. It is clear that the
Shannon entropy is a useful tool in stochastic thermodynamics, and
in thermodynamics of small systems \cite{Seifert2012StochasticReview}.
However, the Shannon entropy (or the von Neumann entropy) is not directly
measurable. There is no Hermitian operator that corresponds to the
entropy. In fact, the entropy is not even linear in the probabilities.
This implies that in order to measure it, the probability distribution
has to be measured via tomography. 

Fortunately, in thermodynamics the standard CI provides a priceless
connection between heat and entropy. For reversible, single-bath processes,
the change in the entropy is given by $\Delta S_{sys}=Q/T$. Moreover,
for irreversible isochores in the weak system bath coupling it holds
that $\Delta S_{bath}=-Q/T_{bath}$ even when $\Delta S_{sys}\gg Q/T_{bath}$.
In the $\alpha$CI the exact same thing holds. We can learn about
the changes in $\Delta\mc S_{\alpha}$ from $\mc Q_{\alpha}/T^{\alpha}$.
\textit{The reason why entropy is important is primarily because it
can be related to heat,} \textit{and secondly because we have some
intuitive understanding of what are high and low entropy states}.
This however is true for any $\alpha$, not just $\alpha=1$. We conclude
that in the context of the present paper $\mc S_{\alpha}$ is just
as useful (or ``physical'') as the regular Shannon entropy used
in thermodynamics of small systems. 

\section*{Appendix VI - An example of using coherence to extract higher order
heat from a bath}

We start this Appendix with a description of an irreversible protocol
for extracting heat from a single bath using coherence erasure (the
protocol in \cite{Anders2015MeasurementWork} is reversible). 

In step A of the protocol a pulse (unitary operation) is applied to
bring the system into a passive state (no coherence and no population
inversion in the energy basis). The new probabilities in the energy
basis are $p_{i}'$. In step B we change the energy level $E_{i}$
without interacting with the bath to $E_{i}=-T\log p_{i}$. In step
C a full isochoric thermalization take place so the density matrix
is equal to $\rho_{\Lambda}$ . Finally, in step D the bath is disconnected
and the levels are adiabatically restored to their original value
$E_{i}$. 

In this protocol $\mc Q_{\alpha}<T^{\alpha}D_{\alpha}^{B}(\rho,\rho_{\Lambda})$.
Since the thermal interaction in this protocol protocol is an isochore,
it is easy to relate it to changes in the bath moments (see Appendix
III). As a concrete example we consider a qutrit system that is initially
in a state
\[
\rho_{s,0}=\left(\begin{array}{ccc}
\nicefrac{1}{6} & \nicefrac{1}{400} & 0\\
\nicefrac{1}{400} & \nicefrac{1}{3} & \nicefrac{1}{20}\\
0 & \nicefrac{1}{20} & \nicefrac{1}{2}
\end{array}\right).
\]
The reason for choosing small coherence values is that we want start
close to thermal equilibrium in order to show that the higher order
bound (\ref{eq: Qa max coh}) can produce a reasonably tight bound.
The diagonals correspond to thermal distribution with temperature
of 1 and a Hamiltonian $H=diagonal[\{\ln3,\ln\tfrac{3}{2},0\}]$.
Since the populations are already in thermal form (Gibbs state) the
irreversible protocol described above, has only two stages: a unitary
rotation pulse (stage A), and an isochore (stage C). 

For stage A we apply an interaction Hamiltonian (in the interaction
picture) 
\[
H_{int}=\left(\begin{array}{ccc}
0 & +i & 0\\
-i & 0 & -i\\
0 & +i & 0
\end{array}\right),
\]
and find that at some point in time ($t_{f}\simeq0.204$), $\left\langle H\right\rangle _{t_{f}}=\left\langle H\right\rangle _{t_{0}}$.
At time $t_{f}$ we start stage C (an isochore). As a result, there
is no heat exchange with the bath. Since the $\mc Q_{1}=0$, $\mc Q_{2}$
expresses the change in energy variance. In this example $\mc Q_{2}\ge0$
which implies, under the condition in Appendix III, that the energy
variance of the particles in the bath is decreasing.

The bound (\ref{eq: Qa max coh}) on the change in variance yields
a value that is 1.92 times large than the actual change in the energy
variance.

\section*{Appendix VII - The low temperature limit}

The goal of this appendix is to explain and show analytically why
the GCI prediction for cold temperature and $\alpha\to0$, converges
to the exact efficiency of the Otto machine (Sec. \ref{subsec: Tighter-than-2nd}
and dashed curve in Fig. \ref{fig: Otto TLS}b). For hot isochore
of the Otto machine the GCI reads:
\begin{align}
\Delta\mc S_{\alpha}^{h}-\beta_{h}^{\alpha}\mc Q_{\alpha}^{h} & \ge0\label{eq: GCI app 7}
\end{align}
Let us write the initial state as
\[
\vec{p}_{c}=\vec{p}_{h}-d\vec{p}
\]
In full thermalization isochores, there are different scenarios in
which $d\vec{p}$ can be small. For example, close to the crossover
to refrigerator $T_{c}/T_{h}=\Delta E^{c}/\Delta E^{h}$ where the
machine becomes reversible, $d\vec{p}$ is very small (at the crossover
the Otto machine satisfies $d\vec{p}=0$ and it produces zero work
per cycle, e.g. see \cite{RUswap}). Another scenario takes place
when $T_{c}\ll\Delta E^{c},T_{h}\ll\Delta E^{h}$. Most of the population
is in the ground state and $\vec{p}_{c}$ and $\vec{p}_{h}$ differ
by a very small number $\exp(-\beta_{c}\Delta E^{c})-\exp(-\beta_{h}\Delta E^{h})$.
However, $T_{c}$ and $T_{h}$ are very different from each other
(not linear response) and very different from the refrigerator crossover
point $T_{c}/T_{h}\neq\Delta E^{c}/\Delta E^{h}$. Since the Carnot
bound also converge to the actual efficiency near the refrigerator
crossover (first scenario), we are interested here only in the second
scenario where both temperatures are low but still very different
from each other or from the refrigerator crossover ratio. Expanding
both terms in (\ref{eq: GCI app 7}) in powers of $\alpha$ and $d\vec{p}$
yields
\begin{align}
\Delta\mc S_{\alpha}^{h} & =\alpha\sum_{j}dp_{j}\ln[-\ln[p_{h,j}]]+O(dp_{j}^{2},\alpha^{2}),\label{eq: dSa series}\\
\beta_{h}^{\alpha}\mc Q_{\alpha}^{h} & =\alpha\sum_{j}dp_{j}\ln[-\ln[p_{h,j}]]+O(dp_{j}^{2},\alpha^{2}).\label{eq: bQa series}
\end{align}
The higher order terms $O(dp_{j}^{2},\alpha^{2})$ in (\ref{eq: dSa series})
and (\ref{eq: bQa series}) differ from each other. From (\ref{eq: dSa series})
and (\ref{eq: bQa series}) we conclude that when both $dp_{j}$ and
$\alpha$ are small, the two terms in (\ref{eq: GCI app 7}) cancel
each other in the lowest order of $\alpha$ and $d\vec{p}$. Thus,
the $\alpha$CI holds as an equality in the limit $d\vec{p}\ll1,\alpha\ll1$.
This explains why the $\alpha$CI prediction (dashed-blue line in
Fig. 3) converges to the actual efficiency although irreversible processes
(isochores) are involved. We point out that on top of low temperatures
$\alpha\ll1$ is also required for (\ref{eq: GCI app 7}) to become
equality. For large $\alpha$ the $O(\alpha^{2})$ terms become important
and the $\alpha$CI is no longer tight. 

\section*{Appendix VIII - choice of parameters for a half-zero heat machine}

In this appendix we describe how to choose the energy levels of the
machine in Fig. \ref{fig: Zero heat machine} in order to achieve
$\mc Q_{1,c}=0$. The cold levels can be chosen freely and we set
them to be $E_{c}=\{0,1,2\}$. For simplicity, it is assumed that
the baths are connected for a period which exceeds several thermalization
times. Hence, at the end of stroke I the populations are $p_{c,j}=e^{-\beta_{c}(E_{c,j}-F_{c})}$.
The cold and hot bath temperatures are $\{T_{c},T_{h}\}=\{0.5,1\}$.
To determine the hot levels, we first choose what is the distribution
$\vec{p}_{h}$ we want the hot bath to generate. To achieve $\mc Q_{1,c}=(\vec{p}_{h}-\vec{p}_{c})\cdot\vec{E}_{c}=0$
we set $\vec{p}_{h}=\vec{p}_{c}+\{-1,2,-1\}\delta p$ where $\delta p$
is taken to be $\delta p=p_{c}(3)/20$. Consequently, the hot energy
levels are $\vec{\mc E}{}_{h}=\vec{E}_{h}-F_{h}=-T_{h}\ln\vec{p}_{h}$.
Choosing one of the levels will fix $F_{h}$ and the values of all
the hot levels. By setting $E_{h,1}=0$ we get $E_{h}\simeq\{0,1.986,4.05\}.$
Nevertheless, for calculating heat (of any order) only $\vec{\mc E}{}_{h}$
is needed. 

To obtain a full zero machine with $Q_{c}=Q_{h}=0$ we choose two
states $\vec{p}_{A,}\vec{p}_{B}$ that have the same Shannon entropy
(at least three levels are needed. See Fig. \ref{fig: S curves} for
an example). A reversible state preparation is used to prepare $\vec{p}_{B}$
from $\vec{p}_{A}$ using a bath in temperature $T_{c}$, and another
reversible state preparation is used to create $\vec{p}_{A}$ from
$\vec{p}_{B}$ using a bath in temperature $T_{h}$.

\bibliographystyle{apsrev4-1}
\bibliography{/Users/raam_uzdin/Dropbox/RaamCite}

\begin{thebibliography}{57}%
\makeatletter
\providecommand \@ifxundefined [1]{%
 \@ifx{#1\undefined}
}%
\providecommand \@ifnum [1]{%
 \ifnum #1\expandafter \@firstoftwo
 \else \expandafter \@secondoftwo
 \fi
}%
\providecommand \@ifx [1]{%
 \ifx #1\expandafter \@firstoftwo
 \else \expandafter \@secondoftwo
 \fi
}%
\providecommand \natexlab [1]{#1}%
\providecommand \enquote  [1]{``#1''}%
\providecommand \bibnamefont  [1]{#1}%
\providecommand \bibfnamefont [1]{#1}%
\providecommand \citenamefont [1]{#1}%
\providecommand \href@noop [0]{\@secondoftwo}%
\providecommand \href [0]{\begingroup \@sanitize@url \@href}%
\providecommand \@href[1]{\@@startlink{#1}\@@href}%
\providecommand \@@href[1]{\endgroup#1\@@endlink}%
\providecommand \@sanitize@url [0]{\catcode `\\12\catcode `\$12\catcode
  `\&12\catcode `\#12\catcode `\^12\catcode `\_12\catcode `\%12\relax}%
\providecommand \@@startlink[1]{}%
\providecommand \@@endlink[0]{}%
\providecommand \url  [0]{\begingroup\@sanitize@url \@url }%
\providecommand \@url [1]{\endgroup\@href {#1}{\urlprefix }}%
\providecommand \urlprefix  [0]{URL }%
\providecommand \Eprint [0]{\href }%
\providecommand \doibase [0]{http://dx.doi.org/}%
\providecommand \selectlanguage [0]{\@gobble}%
\providecommand \bibinfo  [0]{\@secondoftwo}%
\providecommand \bibfield  [0]{\@secondoftwo}%
\providecommand \translation [1]{[#1]}%
\providecommand \BibitemOpen [0]{}%
\providecommand \bibitemStop [0]{}%
\providecommand \bibitemNoStop [0]{.\EOS\space}%
\providecommand \EOS [0]{\spacefactor3000\relax}%
\providecommand \BibitemShut  [1]{\csname bibitem#1\endcsname}%
\let\auto@bib@innerbib\@empty
\bibitem [{\citenamefont {Alicki}(1979)}]{alicki79}%
  \BibitemOpen
  \bibfield  {author} {\bibinfo {author} {\bibfnamefont {R.}~\bibnamefont
  {Alicki}},\ }\href@noop {} {\bibfield  {journal} {\bibinfo  {journal} {J.
  Phys A: Math.Gen.}\ }\textbf {\bibinfo {volume} {12}},\ \bibinfo {pages}
  {L103} (\bibinfo {year} {1979})}\BibitemShut {NoStop}%
\bibitem [{\citenamefont {Sagawa}(2012)}]{Sagawa2012second}%
  \BibitemOpen
  \bibfield  {author} {\bibinfo {author} {\bibfnamefont {T.}~\bibnamefont
  {Sagawa}},\ }\href@noop {} {\bibfield  {journal} {\bibinfo  {journal}
  {Lectures on Quantum Computing, Thermodynamics and Statistical Physics}\
  }\textbf {\bibinfo {volume} {8}},\ \bibinfo {pages} {127} (\bibinfo {year}
  {2012})}\BibitemShut {NoStop}%
\bibitem [{\citenamefont {Peres}(2006)}]{PeresBook}%
  \BibitemOpen
  \bibfield  {author} {\bibinfo {author} {\bibfnamefont {A.}~\bibnamefont
  {Peres}},\ }\href@noop {} {\emph {\bibinfo {title} {Quantum theory: concepts
  and methods}}},\ Vol.~\bibinfo {volume} {57}\ (\bibinfo  {publisher}
  {Springer Science \& Business Media},\ \bibinfo {year} {2006})\BibitemShut
  {NoStop}%
\bibitem [{\citenamefont {Esposito}\ and\ \citenamefont {Van~den
  Broeck}(2011)}]{Esposito2011EPL2Law}%
  \BibitemOpen
  \bibfield  {author} {\bibinfo {author} {\bibfnamefont {M.}~\bibnamefont
  {Esposito}}\ and\ \bibinfo {author} {\bibfnamefont {C.}~\bibnamefont {Van~den
  Broeck}},\ }\href@noop {} {\bibfield  {journal} {\bibinfo  {journal} {EPL
  (Europhysics Letters)}\ }\textbf {\bibinfo {volume} {95}},\ \bibinfo {pages}
  {40004} (\bibinfo {year} {2011})}\BibitemShut {NoStop}%
\bibitem [{\citenamefont {Uzdin}\ \emph {et~al.}(2015)\citenamefont {Uzdin},
  \citenamefont {Levy},\ and\ \citenamefont {Kosloff}}]{EquivPRX}%
  \BibitemOpen
  \bibfield  {author} {\bibinfo {author} {\bibfnamefont {R.}~\bibnamefont
  {Uzdin}}, \bibinfo {author} {\bibfnamefont {A.}~\bibnamefont {Levy}}, \ and\
  \bibinfo {author} {\bibfnamefont {R.}~\bibnamefont {Kosloff}},\ }\href@noop
  {} {\bibfield  {journal} {\bibinfo  {journal} {Phys. Rev. X}\ }\textbf
  {\bibinfo {volume} {5}},\ \bibinfo {pages} {031044} (\bibinfo {year}
  {2015})}\BibitemShut {NoStop}%
\bibitem [{\citenamefont {Gelbwaser-Klimovsky}\ and\ \citenamefont
  {Kurizki}(2014)}]{gelbwaser2014heat}%
  \BibitemOpen
  \bibfield  {author} {\bibinfo {author} {\bibfnamefont {D.}~\bibnamefont
  {Gelbwaser-Klimovsky}}\ and\ \bibinfo {author} {\bibfnamefont
  {G.}~\bibnamefont {Kurizki}},\ }\href@noop {} {\bibfield  {journal} {\bibinfo
   {journal} {Physical Review E}\ }\textbf {\bibinfo {volume} {90}},\ \bibinfo
  {pages} {022102} (\bibinfo {year} {2014})}\BibitemShut {NoStop}%
\bibitem [{\citenamefont {Mitchison}\ \emph {et~al.}(2015)\citenamefont
  {Mitchison}, \citenamefont {Woods}, \citenamefont {Prior},\ and\
  \citenamefont {Huber}}]{MitchisonHuber2015CoherenceAssitedCooling}%
  \BibitemOpen
  \bibfield  {author} {\bibinfo {author} {\bibfnamefont {M.~T.}\ \bibnamefont
  {Mitchison}}, \bibinfo {author} {\bibfnamefont {M.~P.}\ \bibnamefont
  {Woods}}, \bibinfo {author} {\bibfnamefont {J.}~\bibnamefont {Prior}}, \ and\
  \bibinfo {author} {\bibfnamefont {M.}~\bibnamefont {Huber}},\ }\href@noop {}
  {\bibfield  {journal} {\bibinfo  {journal} {New Journal of Physics}\ }\textbf
  {\bibinfo {volume} {17}},\ \bibinfo {pages} {115013} (\bibinfo {year}
  {2015})}\BibitemShut {NoStop}%
\bibitem [{\citenamefont {Andrieux}\ and\ \citenamefont
  {Gaspard}(2008)}]{GaspardCopyDNA}%
  \BibitemOpen
  \bibfield  {author} {\bibinfo {author} {\bibfnamefont {D.}~\bibnamefont
  {Andrieux}}\ and\ \bibinfo {author} {\bibfnamefont {P.}~\bibnamefont
  {Gaspard}},\ }\href@noop {} {\bibfield  {journal} {\bibinfo  {journal}
  {Proceedings of the National Academy of Sciences}\ }\textbf {\bibinfo
  {volume} {105}},\ \bibinfo {pages} {9516} (\bibinfo {year}
  {2008})}\BibitemShut {NoStop}%
\bibitem [{\citenamefont {Jarzynski}(2008)}]{JarzynskiOnGaspard}%
  \BibitemOpen
  \bibfield  {author} {\bibinfo {author} {\bibfnamefont {C.}~\bibnamefont
  {Jarzynski}},\ }\href@noop {} {\bibfield  {journal} {\bibinfo  {journal}
  {Proceedings of the National Academy of Sciences}\ }\textbf {\bibinfo
  {volume} {105}},\ \bibinfo {pages} {9451} (\bibinfo {year}
  {2008})}\BibitemShut {NoStop}%
\bibitem [{\citenamefont {Seifert}(2012)}]{Seifert2012StochasticReview}%
  \BibitemOpen
  \bibfield  {author} {\bibinfo {author} {\bibfnamefont {U.}~\bibnamefont
  {Seifert}},\ }\href@noop {} {\bibfield  {journal} {\bibinfo  {journal}
  {Reports on Progress in Physics}\ }\textbf {\bibinfo {volume} {75}},\
  \bibinfo {pages} {126001} (\bibinfo {year} {2012})}\BibitemShut {NoStop}%
\bibitem [{\citenamefont {Harris}\ and\ \citenamefont
  {Sch{\"u}tz}(2007)}]{harris2007fluctuationReview}%
  \BibitemOpen
  \bibfield  {author} {\bibinfo {author} {\bibfnamefont {R.}~\bibnamefont
  {Harris}}\ and\ \bibinfo {author} {\bibfnamefont {G.}~\bibnamefont
  {Sch{\"u}tz}},\ }\href@noop {} {\bibfield  {journal} {\bibinfo  {journal}
  {Journal of Statistical Mechanics: Theory and Experiment}\ }\textbf {\bibinfo
  {volume} {2007}},\ \bibinfo {pages} {P07020} (\bibinfo {year}
  {2007})}\BibitemShut {NoStop}%
\bibitem [{\citenamefont {Jarzynski}(2011)}]{Jarzynski2011equalitiesReview}%
  \BibitemOpen
  \bibfield  {author} {\bibinfo {author} {\bibfnamefont {C.}~\bibnamefont
  {Jarzynski}},\ }\href@noop {} {\bibfield  {journal} {\bibinfo  {journal}
  {Annu. Rev. Condens. Matter Phys.}\ }\textbf {\bibinfo {volume} {2}},\
  \bibinfo {pages} {329} (\bibinfo {year} {2011})}\BibitemShut {NoStop}%
\bibitem [{\citenamefont {Ro{\ss}nagel}\ \emph {et~al.}(2016)\citenamefont
  {Ro{\ss}nagel}, \citenamefont {Dawkins}, \citenamefont {Tolazzi},
  \citenamefont {Abah}, \citenamefont {Lutz}, \citenamefont {Schmidt-Kaler},\
  and\ \citenamefont {Singer}}]{rossnagelIonEngExp}%
  \BibitemOpen
  \bibfield  {author} {\bibinfo {author} {\bibfnamefont {J.}~\bibnamefont
  {Ro{\ss}nagel}}, \bibinfo {author} {\bibfnamefont {S.~T.}\ \bibnamefont
  {Dawkins}}, \bibinfo {author} {\bibfnamefont {K.~N.}\ \bibnamefont
  {Tolazzi}}, \bibinfo {author} {\bibfnamefont {O.}~\bibnamefont {Abah}},
  \bibinfo {author} {\bibfnamefont {E.}~\bibnamefont {Lutz}}, \bibinfo {author}
  {\bibfnamefont {F.}~\bibnamefont {Schmidt-Kaler}}, \ and\ \bibinfo {author}
  {\bibfnamefont {K.}~\bibnamefont {Singer}},\ }\href@noop {} {\bibfield
  {journal} {\bibinfo  {journal} {Science}\ }\textbf {\bibinfo {volume}
  {352}},\ \bibinfo {pages} {325} (\bibinfo {year} {2016})}\BibitemShut
  {NoStop}%
\bibitem [{\citenamefont {Maslennikov}\ \emph {et~al.}(2017)\citenamefont
  {Maslennikov}, \citenamefont {Ding}, \citenamefont {Hablutzel}, \citenamefont
  {Gan}, \citenamefont {Roulet}, \citenamefont {Nimmrichter}, \citenamefont
  {Dai}, \citenamefont {Scarani},\ and\ \citenamefont
  {Matsukevich}}]{Sing3ionEng2017}%
  \BibitemOpen
  \bibfield  {author} {\bibinfo {author} {\bibfnamefont {G.}~\bibnamefont
  {Maslennikov}}, \bibinfo {author} {\bibfnamefont {S.}~\bibnamefont {Ding}},
  \bibinfo {author} {\bibfnamefont {R.}~\bibnamefont {Hablutzel}}, \bibinfo
  {author} {\bibfnamefont {J.}~\bibnamefont {Gan}}, \bibinfo {author}
  {\bibfnamefont {A.}~\bibnamefont {Roulet}}, \bibinfo {author} {\bibfnamefont
  {S.}~\bibnamefont {Nimmrichter}}, \bibinfo {author} {\bibfnamefont
  {J.}~\bibnamefont {Dai}}, \bibinfo {author} {\bibfnamefont {V.}~\bibnamefont
  {Scarani}}, \ and\ \bibinfo {author} {\bibfnamefont {D.}~\bibnamefont
  {Matsukevich}},\ }\href@noop {} {\bibfield  {journal} {\bibinfo  {journal}
  {arXiv preprint arXiv:1702.08672}\ } (\bibinfo {year} {2017})}\BibitemShut
  {NoStop}%
\bibitem [{\citenamefont {Niskanen}\ \emph {et~al.}(2007)\citenamefont
  {Niskanen}, \citenamefont {Nakamura},\ and\ \citenamefont
  {Pekola}}]{PekolaSCengine}%
  \BibitemOpen
  \bibfield  {author} {\bibinfo {author} {\bibfnamefont {A.}~\bibnamefont
  {Niskanen}}, \bibinfo {author} {\bibfnamefont {Y.}~\bibnamefont {Nakamura}},
  \ and\ \bibinfo {author} {\bibfnamefont {J.}~\bibnamefont {Pekola}},\
  }\href@noop {} {\bibfield  {journal} {\bibinfo  {journal} {Physical Review
  B}\ }\textbf {\bibinfo {volume} {76}},\ \bibinfo {pages} {174523} (\bibinfo
  {year} {2007})}\BibitemShut {NoStop}%
\bibitem [{\citenamefont {Campisi}\ \emph {et~al.}(2015)\citenamefont
  {Campisi}, \citenamefont {Pekola},\ and\ \citenamefont
  {Fazio}}]{campisi2014FT_SolidStateExp}%
  \BibitemOpen
  \bibfield  {author} {\bibinfo {author} {\bibfnamefont {M.}~\bibnamefont
  {Campisi}}, \bibinfo {author} {\bibfnamefont {J.}~\bibnamefont {Pekola}}, \
  and\ \bibinfo {author} {\bibfnamefont {R.}~\bibnamefont {Fazio}},\
  }\href@noop {} {\bibfield  {journal} {\bibinfo  {journal} {New Journal of
  Physics}\ }\textbf {\bibinfo {volume} {17}},\ \bibinfo {pages} {035012}
  (\bibinfo {year} {2015})}\BibitemShut {NoStop}%
\bibitem [{\citenamefont {Gelbwaser-Klimovsky}\ and\ \citenamefont
  {Kurizki}(2015)}]{Kurizki2015workOptoMech}%
  \BibitemOpen
  \bibfield  {author} {\bibinfo {author} {\bibfnamefont {D.}~\bibnamefont
  {Gelbwaser-Klimovsky}}\ and\ \bibinfo {author} {\bibfnamefont
  {G.}~\bibnamefont {Kurizki}},\ }\href@noop {} {\bibfield  {journal} {\bibinfo
   {journal} {Scientific reports}\ }\textbf {\bibinfo {volume} {5}} (\bibinfo
  {year} {2015})}\BibitemShut {NoStop}%
\bibitem [{\citenamefont {Zhang}\ \emph {et~al.}(2014)\citenamefont {Zhang},
  \citenamefont {Bariani},\ and\ \citenamefont {Meystre}}]{ZhangOptoMechEng}%
  \BibitemOpen
  \bibfield  {author} {\bibinfo {author} {\bibfnamefont {K.}~\bibnamefont
  {Zhang}}, \bibinfo {author} {\bibfnamefont {F.}~\bibnamefont {Bariani}}, \
  and\ \bibinfo {author} {\bibfnamefont {P.}~\bibnamefont {Meystre}},\
  }\href@noop {} {\bibfield  {journal} {\bibinfo  {journal} {Physical review
  letters}\ }\textbf {\bibinfo {volume} {112}},\ \bibinfo {pages} {150602}
  (\bibinfo {year} {2014})}\BibitemShut {NoStop}%
\bibitem [{\citenamefont {Mitchison}\ \emph {et~al.}(2016)\citenamefont
  {Mitchison}, \citenamefont {Huber}, \citenamefont {Prior}, \citenamefont
  {Woods},\ and\ \citenamefont {Plenio}}]{Mitchison2016cavityQED}%
  \BibitemOpen
  \bibfield  {author} {\bibinfo {author} {\bibfnamefont {M.~T.}\ \bibnamefont
  {Mitchison}}, \bibinfo {author} {\bibfnamefont {M.}~\bibnamefont {Huber}},
  \bibinfo {author} {\bibfnamefont {J.}~\bibnamefont {Prior}}, \bibinfo
  {author} {\bibfnamefont {M.~P.}\ \bibnamefont {Woods}}, \ and\ \bibinfo
  {author} {\bibfnamefont {M.~B.}\ \bibnamefont {Plenio}},\ }\href@noop {}
  {\bibfield  {journal} {\bibinfo  {journal} {Quantum Science and Technology}\
  }\textbf {\bibinfo {volume} {1}},\ \bibinfo {pages} {015001} (\bibinfo {year}
  {2016})}\BibitemShut {NoStop}%
\bibitem [{\citenamefont {Jarzynski}(1999)}]{JarzynskiMicroscopicClausius}%
  \BibitemOpen
  \bibfield  {author} {\bibinfo {author} {\bibfnamefont {C.}~\bibnamefont
  {Jarzynski}},\ }\href {\doibase 10.1023/A:1004541004050} {\bibfield
  {journal} {\bibinfo  {journal} {Journal of Statistical Physics}\ }\textbf
  {\bibinfo {volume} {96}},\ \bibinfo {pages} {415} (\bibinfo {year}
  {1999})}\BibitemShut {NoStop}%
\bibitem [{\citenamefont {Quan}\ and\ \citenamefont
  {Dong}(2008)}]{quan2008quantumFluctTheorem}%
  \BibitemOpen
  \bibfield  {author} {\bibinfo {author} {\bibfnamefont {H.}~\bibnamefont
  {Quan}}\ and\ \bibinfo {author} {\bibfnamefont {H.}~\bibnamefont {Dong}},\
  }\href@noop {} {\bibfield  {journal} {\bibinfo  {journal} {arXiv preprint
  arXiv:0812.4955}\ } (\bibinfo {year} {2008})}\BibitemShut {NoStop}%
\bibitem [{\citenamefont {Campisi}(2014)}]{campisi14}%
  \BibitemOpen
  \bibfield  {author} {\bibinfo {author} {\bibfnamefont {M.}~\bibnamefont
  {Campisi}},\ }\href@noop {} {\bibfield  {journal} {\bibinfo  {journal} {J.
  Phys A: Math.theor.}\ }\textbf {\bibinfo {volume} {{47}}},\ \bibinfo {pages}
  {{245001}} (\bibinfo {year} {{2014}})}\BibitemShut {NoStop}%
\bibitem [{\citenamefont {Goold}\ \emph {et~al.}(2016)\citenamefont {Goold},
  \citenamefont {Huber}, \citenamefont {Riera}, \citenamefont {del Rio},\ and\
  \citenamefont {Skrzypczyk}}]{Goold2015review}%
  \BibitemOpen
  \bibfield  {author} {\bibinfo {author} {\bibfnamefont {J.}~\bibnamefont
  {Goold}}, \bibinfo {author} {\bibfnamefont {M.}~\bibnamefont {Huber}},
  \bibinfo {author} {\bibfnamefont {A.}~\bibnamefont {Riera}}, \bibinfo
  {author} {\bibfnamefont {L.}~\bibnamefont {del Rio}}, \ and\ \bibinfo
  {author} {\bibfnamefont {P.}~\bibnamefont {Skrzypczyk}},\ }\href@noop {}
  {\bibfield  {journal} {\bibinfo  {journal} {Journal of Physics A:
  Mathematical and Theoretical}\ }\textbf {\bibinfo {volume} {49}},\ \bibinfo
  {pages} {143001} (\bibinfo {year} {2016})}\BibitemShut {NoStop}%
\bibitem [{\citenamefont {Horodecki}\ and\ \citenamefont
  {Oppenheim}(2013)}]{horodecki2013fundamental}%
  \BibitemOpen
  \bibfield  {author} {\bibinfo {author} {\bibfnamefont {M.}~\bibnamefont
  {Horodecki}}\ and\ \bibinfo {author} {\bibfnamefont {J.}~\bibnamefont
  {Oppenheim}},\ }\href@noop {} {\bibfield  {journal} {\bibinfo  {journal}
  {Nature communications}\ }\textbf {\bibinfo {volume} {4}},\ \bibinfo {pages}
  {2059} (\bibinfo {year} {2013})}\BibitemShut {NoStop}%
\bibitem [{\citenamefont {Gour}\ \emph {et~al.}(2015)\citenamefont {Gour},
  \citenamefont {M{\"u}ller}, \citenamefont {Narasimhachar}, \citenamefont
  {Spekkens},\ and\ \citenamefont {Halpern}}]{GourRTreview}%
  \BibitemOpen
  \bibfield  {author} {\bibinfo {author} {\bibfnamefont {G.}~\bibnamefont
  {Gour}}, \bibinfo {author} {\bibfnamefont {M.~P.}\ \bibnamefont
  {M{\"u}ller}}, \bibinfo {author} {\bibfnamefont {V.}~\bibnamefont
  {Narasimhachar}}, \bibinfo {author} {\bibfnamefont {R.~W.}\ \bibnamefont
  {Spekkens}}, \ and\ \bibinfo {author} {\bibfnamefont {N.~Y.}\ \bibnamefont
  {Halpern}},\ }\href@noop {} {\bibfield  {journal} {\bibinfo  {journal}
  {Physics Reports}\ }\textbf {\bibinfo {volume} {583}},\ \bibinfo {pages} {1}
  (\bibinfo {year} {2015})}\BibitemShut {NoStop}%
\bibitem [{\citenamefont {Lostaglio}\ \emph
  {et~al.}(2015{\natexlab{a}})\citenamefont {Lostaglio}, \citenamefont
  {Jennings},\ and\ \citenamefont {Rudolph}}]{LostaglioRudolphCohConstraint}%
  \BibitemOpen
  \bibfield  {author} {\bibinfo {author} {\bibfnamefont {M.}~\bibnamefont
  {Lostaglio}}, \bibinfo {author} {\bibfnamefont {D.}~\bibnamefont {Jennings}},
  \ and\ \bibinfo {author} {\bibfnamefont {T.}~\bibnamefont {Rudolph}},\
  }\href@noop {} {\bibfield  {journal} {\bibinfo  {journal} {Nature
  communications}\ }\textbf {\bibinfo {volume} {6}},\ \bibinfo {pages} {6383}
  (\bibinfo {year} {2015}{\natexlab{a}})}\BibitemShut {NoStop}%
\bibitem [{\citenamefont {Lostaglio}\ \emph
  {et~al.}(2015{\natexlab{b}})\citenamefont {Lostaglio}, \citenamefont
  {M\"uller},\ and\ \citenamefont {Pastena}}]{MatteoIndependenceResource}%
  \BibitemOpen
  \bibfield  {author} {\bibinfo {author} {\bibfnamefont {M.}~\bibnamefont
  {Lostaglio}}, \bibinfo {author} {\bibfnamefont {M.~P.}\ \bibnamefont
  {M\"uller}}, \ and\ \bibinfo {author} {\bibfnamefont {M.}~\bibnamefont
  {Pastena}},\ }\href {\doibase 10.1103/PhysRevLett.115.150402} {\bibfield
  {journal} {\bibinfo  {journal} {Phys. Rev. Lett.}\ }\textbf {\bibinfo
  {volume} {115}},\ \bibinfo {pages} {150402} (\bibinfo {year}
  {2015}{\natexlab{b}})}\BibitemShut {NoStop}%
\bibitem [{\citenamefont {Deffner}\ and\ \citenamefont
  {Lutz}(2010)}]{DeffnerLutzRelEntBures}%
  \BibitemOpen
  \bibfield  {author} {\bibinfo {author} {\bibfnamefont {S.}~\bibnamefont
  {Deffner}}\ and\ \bibinfo {author} {\bibfnamefont {E.}~\bibnamefont {Lutz}},\
  }\href@noop {} {\bibfield  {journal} {\bibinfo  {journal} {Physical review
  letters}\ }\textbf {\bibinfo {volume} {105}},\ \bibinfo {pages} {170402}
  (\bibinfo {year} {2010})}\BibitemShut {NoStop}%
\bibitem [{\citenamefont {Anders}\ and\ \citenamefont
  {Giovannetti}(2013)}]{anders2013thermodynamics}%
  \BibitemOpen
  \bibfield  {author} {\bibinfo {author} {\bibfnamefont {J.}~\bibnamefont
  {Anders}}\ and\ \bibinfo {author} {\bibfnamefont {V.}~\bibnamefont
  {Giovannetti}},\ }\href@noop {} {\bibfield  {journal} {\bibinfo  {journal}
  {New Journal of Physics}\ }\textbf {\bibinfo {volume} {15}},\ \bibinfo
  {pages} {033022} (\bibinfo {year} {2013})}\BibitemShut {NoStop}%
\bibitem [{\citenamefont {Bregman}(1967)}]{Bregman1967}%
  \BibitemOpen
  \bibfield  {author} {\bibinfo {author} {\bibfnamefont {L.~M.}\ \bibnamefont
  {Bregman}},\ }\href@noop {} {\bibfield  {journal} {\bibinfo  {journal} {USSR
  computational mathematics and mathematical physics}\ }\textbf {\bibinfo
  {volume} {7}},\ \bibinfo {pages} {200} (\bibinfo {year} {1967})}\BibitemShut
  {NoStop}%
\bibitem [{\citenamefont {Still}\ \emph {et~al.}(2012)\citenamefont {Still},
  \citenamefont {Sivak}, \citenamefont {Bell},\ and\ \citenamefont
  {Crooks}}]{CrooksThemoPredNonEf}%
  \BibitemOpen
  \bibfield  {author} {\bibinfo {author} {\bibfnamefont {S.}~\bibnamefont
  {Still}}, \bibinfo {author} {\bibfnamefont {D.~A.}\ \bibnamefont {Sivak}},
  \bibinfo {author} {\bibfnamefont {A.~J.}\ \bibnamefont {Bell}}, \ and\
  \bibinfo {author} {\bibfnamefont {G.~E.}\ \bibnamefont {Crooks}},\
  }\href@noop {} {\bibfield  {journal} {\bibinfo  {journal} {Physical review
  letters}\ }\textbf {\bibinfo {volume} {109}},\ \bibinfo {pages} {120604}
  (\bibinfo {year} {2012})}\BibitemShut {NoStop}%
\bibitem [{\citenamefont {Reeb}\ and\ \citenamefont
  {Wolf}(2014)}]{reeb2014improved}%
  \BibitemOpen
  \bibfield  {author} {\bibinfo {author} {\bibfnamefont {D.}~\bibnamefont
  {Reeb}}\ and\ \bibinfo {author} {\bibfnamefont {M.~M.}\ \bibnamefont
  {Wolf}},\ }\href@noop {} {\bibfield  {journal} {\bibinfo  {journal} {New
  Journal of Physics}\ }\textbf {\bibinfo {volume} {16}},\ \bibinfo {pages}
  {103011} (\bibinfo {year} {2014})}\BibitemShut {NoStop}%
\bibitem [{Note1()}]{Note1}%
  \BibitemOpen
  \bibinfo {note} {Follow immediately from the convexity of $-\protect \mathcal
  {S}$ and the fact that a linear term does not affect concavity.}\BibitemShut
  {Stop}%
\bibitem [{\citenamefont {Spohn}(1978)}]{spohn78}%
  \BibitemOpen
  \bibfield  {author} {\bibinfo {author} {\bibfnamefont {H.}~\bibnamefont
  {Spohn}},\ }\href@noop {} {\bibfield  {journal} {\bibinfo  {journal} {Journal
  of Mathematical Physics}\ }\textbf {\bibinfo {volume} {19}},\ \bibinfo
  {pages} {1227} (\bibinfo {year} {1978})}\BibitemShut {NoStop}%
\bibitem [{\citenamefont {{H.-P. Breuer and F. Petruccione}}(2002)}]{breuer}%
  \BibitemOpen
  \bibfield  {author} {\bibinfo {author} {\bibnamefont {{H.-P. Breuer and F.
  Petruccione}}},\ }\href@noop {} {\emph {\bibinfo {title} {Open quantum
  systems}}}\ (\bibinfo  {publisher} {Oxford university press},\ \bibinfo
  {year} {2002})\BibitemShut {NoStop}%
\bibitem [{\citenamefont {{A. E. Allahverdyan, R. Balian, and Th. M.
  Nieuwenhuizen}}(2004)}]{AllahverdyanErgotropy}%
  \BibitemOpen
  \bibfield  {author} {\bibinfo {author} {\bibnamefont {{A. E. Allahverdyan, R.
  Balian, and Th. M. Nieuwenhuizen}}},\ }\href@noop {} {\bibfield  {journal}
  {\bibinfo  {journal} {Euro. Phys. Lett.}\ }\textbf {\bibinfo {volume} {67}},\
  \bibinfo {pages} {{565}} (\bibinfo {year} {2004})}\BibitemShut {NoStop}%
\bibitem [{\citenamefont {Havrda}\ and\ \citenamefont
  {Charv{\'a}t}(1967)}]{havrda1967EarlyTsallisDef}%
  \BibitemOpen
  \bibfield  {author} {\bibinfo {author} {\bibfnamefont {J.}~\bibnamefont
  {Havrda}}\ and\ \bibinfo {author} {\bibfnamefont {F.}~\bibnamefont
  {Charv{\'a}t}},\ }\href@noop {} {\bibfield  {journal} {\bibinfo  {journal}
  {Kybernetika}\ }\textbf {\bibinfo {volume} {3}},\ \bibinfo {pages} {30}
  (\bibinfo {year} {1967})}\BibitemShut {NoStop}%
\bibitem [{\citenamefont {Vajda}(1968)}]{vajda1968EarlyTsallisDef}%
  \BibitemOpen
  \bibfield  {author} {\bibinfo {author} {\bibfnamefont {I.}~\bibnamefont
  {Vajda}},\ }\href@noop {} {\bibfield  {journal} {\bibinfo  {journal}
  {Kybernetika}\ }\textbf {\bibinfo {volume} {4}},\ \bibinfo {pages} {105}
  (\bibinfo {year} {1968})}\BibitemShut {NoStop}%
\bibitem [{\citenamefont {Dar{\'o}czy}(1970)}]{daroczy1970EarlyTsallisDef}%
  \BibitemOpen
  \bibfield  {author} {\bibinfo {author} {\bibfnamefont {Z.}~\bibnamefont
  {Dar{\'o}czy}},\ }\href@noop {} {\bibfield  {journal} {\bibinfo  {journal}
  {Information and control}\ }\textbf {\bibinfo {volume} {16}},\ \bibinfo
  {pages} {36} (\bibinfo {year} {1970})}\BibitemShut {NoStop}%
\bibitem [{\citenamefont {Tsallis}(1988)}]{TsalisOriginal}%
  \BibitemOpen
  \bibfield  {author} {\bibinfo {author} {\bibfnamefont {C.}~\bibnamefont
  {Tsallis}},\ }\href@noop {} {\bibfield  {journal} {\bibinfo  {journal}
  {Journal of statistical physics}\ }\textbf {\bibinfo {volume} {52}},\
  \bibinfo {pages} {479} (\bibinfo {year} {1988})}\BibitemShut {NoStop}%
\bibitem [{\citenamefont {Abe}\ and\ \citenamefont
  {Okamoto}(2001)}]{TsalisStatMech}%
  \BibitemOpen
  \bibfield  {author} {\bibinfo {author} {\bibfnamefont {S.}~\bibnamefont
  {Abe}}\ and\ \bibinfo {author} {\bibfnamefont {Y.}~\bibnamefont {Okamoto}},\
  }\href@noop {} {\emph {\bibinfo {title} {Nonextensive statistical mechanics
  and its applications}}},\ Vol.\ \bibinfo {volume} {560}\ (\bibinfo
  {publisher} {Springer Science \& Business Media},\ \bibinfo {year}
  {2001})\BibitemShut {NoStop}%
\bibitem [{Note2()}]{Note2}%
  \BibitemOpen
  \bibinfo {note} {Purity is defined $tr[\rho ^{2}]$ and we call one minus the
  purity the ``impurity'' of $\rho $. The $\protect \mathaccentV
  {tilde}07E{\alpha }$ impurity is best known as Tsallis entropy. However, the
  physical context of Tsallis entropy is associated with non thermal Tsallis
  distribution. Since our reservoirs are always thermal we use a different name
  to prevent possible confusion.}\BibitemShut {Stop}%
\bibitem [{\citenamefont {Nock}\ \emph {et~al.}(2013)\citenamefont {Nock},
  \citenamefont {Magdalou}, \citenamefont {Briys},\ and\ \citenamefont
  {Nielsen}}]{FNielsen2013miningBregmanMatrix}%
  \BibitemOpen
  \bibfield  {author} {\bibinfo {author} {\bibfnamefont {R.}~\bibnamefont
  {Nock}}, \bibinfo {author} {\bibfnamefont {B.}~\bibnamefont {Magdalou}},
  \bibinfo {author} {\bibfnamefont {E.}~\bibnamefont {Briys}}, \ and\ \bibinfo
  {author} {\bibfnamefont {F.}~\bibnamefont {Nielsen}},\ }in\ \href@noop {}
  {\emph {\bibinfo {booktitle} {Matrix Information Geometry}}}\ (\bibinfo
  {publisher} {Springer},\ \bibinfo {year} {2013})\ pp.\ \bibinfo {pages}
  {373--402}\BibitemShut {NoStop}%
\bibitem [{\citenamefont {Moln{\'a}r}\ \emph {et~al.}(2016)\citenamefont
  {Moln{\'a}r}, \citenamefont {Pitrik},\ and\ \citenamefont
  {Virosztek}}]{molnar2016mapsBregmanMatrix}%
  \BibitemOpen
  \bibfield  {author} {\bibinfo {author} {\bibfnamefont {L.}~\bibnamefont
  {Moln{\'a}r}}, \bibinfo {author} {\bibfnamefont {J.}~\bibnamefont {Pitrik}},
  \ and\ \bibinfo {author} {\bibfnamefont {D.}~\bibnamefont {Virosztek}},\
  }\href@noop {} {\bibfield  {journal} {\bibinfo  {journal} {Linear Algebra and
  its Applications}\ }\textbf {\bibinfo {volume} {495}},\ \bibinfo {pages}
  {174} (\bibinfo {year} {2016})}\BibitemShut {NoStop}%
\bibitem [{Note3()}]{Note3}%
  \BibitemOpen
  \bibinfo {note} {We use the Bregman divergence only for density matrices.
  Hence, conjugation and transposition are not needed.}\BibitemShut {Stop}%
\bibitem [{\citenamefont {Baumgratz}\ \emph {et~al.}(2014)\citenamefont
  {Baumgratz}, \citenamefont {Cramer},\ and\ \citenamefont
  {Plenio}}]{PlenioCoherence}%
  \BibitemOpen
  \bibfield  {author} {\bibinfo {author} {\bibfnamefont {T.}~\bibnamefont
  {Baumgratz}}, \bibinfo {author} {\bibfnamefont {M.}~\bibnamefont {Cramer}}, \
  and\ \bibinfo {author} {\bibfnamefont {M.}~\bibnamefont {Plenio}},\
  }\href@noop {} {\bibfield  {journal} {\bibinfo  {journal} {Physical review
  letters}\ }\textbf {\bibinfo {volume} {113}},\ \bibinfo {pages} {140401}
  (\bibinfo {year} {2014})}\BibitemShut {NoStop}%
\bibitem [{\citenamefont {Kammerlander}\ and\ \citenamefont
  {Anders}(2016)}]{Anders2015MeasurementWork}%
  \BibitemOpen
  \bibfield  {author} {\bibinfo {author} {\bibfnamefont {P.}~\bibnamefont
  {Kammerlander}}\ and\ \bibinfo {author} {\bibfnamefont {J.}~\bibnamefont
  {Anders}},\ }\href@noop {} {\bibfield  {journal} {\bibinfo  {journal}
  {Scientific Reports}\ }\textbf {\bibinfo {volume} {6}},\ \bibinfo {pages}
  {22174} (\bibinfo {year} {2016})}\BibitemShut {NoStop}%
\bibitem [{\citenamefont {Uzdin}\ and\ \citenamefont {Kosloff}(2014)}]{RUswap}%
  \BibitemOpen
  \bibfield  {author} {\bibinfo {author} {\bibfnamefont {R.}~\bibnamefont
  {Uzdin}}\ and\ \bibinfo {author} {\bibfnamefont {R.}~\bibnamefont
  {Kosloff}},\ }\href@noop {} {\bibfield  {journal} {\bibinfo  {journal} {New
  Journal of Physics}\ }\textbf {\bibinfo {volume} {{16}}},\ \bibinfo {pages}
  {{095003}} (\bibinfo {year} {{2014}})}\BibitemShut {NoStop}%
\bibitem [{\citenamefont {Kay}\ and\ \citenamefont
  {Leigh}(2015)}]{Leigh2015RiseMachines}%
  \BibitemOpen
  \bibfield  {author} {\bibinfo {author} {\bibfnamefont {E.~R.}\ \bibnamefont
  {Kay}}\ and\ \bibinfo {author} {\bibfnamefont {D.~A.}\ \bibnamefont
  {Leigh}},\ }\href@noop {} {\bibfield  {journal} {\bibinfo  {journal}
  {Angewandte Chemie International Edition}\ }\textbf {\bibinfo {volume}
  {54}},\ \bibinfo {pages} {10080} (\bibinfo {year} {2015})}\BibitemShut
  {NoStop}%
\bibitem [{\citenamefont {{Eitan Geva and Ronnie Kosloff}}(1994)}]{k102}%
  \BibitemOpen
  \bibfield  {author} {\bibinfo {author} {\bibnamefont {{Eitan Geva and Ronnie
  Kosloff}}},\ }\href@noop {} {\bibfield  {journal} {\bibinfo  {journal} {Phys.
  Rev. E}\ }\textbf {\bibinfo {volume} {49}},\ \bibinfo {pages} {3903}
  (\bibinfo {year} {1994})}\BibitemShut {NoStop}%
\bibitem [{\citenamefont {Scovil}\ and\ \citenamefont
  {Schulz-DuBois}(1959)}]{scovil59}%
  \BibitemOpen
  \bibfield  {author} {\bibinfo {author} {\bibfnamefont {H.~E.~D.}\
  \bibnamefont {Scovil}}\ and\ \bibinfo {author} {\bibfnamefont {E.~O.}\
  \bibnamefont {Schulz-DuBois}},\ }\href@noop {} {\bibfield  {journal}
  {\bibinfo  {journal} {Phys. Rev. Lett.}\ }\textbf {\bibinfo {volume} {2}},\
  \bibinfo {pages} {262} (\bibinfo {year} {1959})}\BibitemShut {NoStop}%
\bibitem [{\citenamefont {Uzdin}\ \emph {et~al.}(2016)\citenamefont {Uzdin},
  \citenamefont {Levy},\ and\ \citenamefont {Kosloff}}]{RUnonMarkovianEquiv}%
  \BibitemOpen
  \bibfield  {author} {\bibinfo {author} {\bibfnamefont {R.}~\bibnamefont
  {Uzdin}}, \bibinfo {author} {\bibfnamefont {A.}~\bibnamefont {Levy}}, \ and\
  \bibinfo {author} {\bibfnamefont {R.}~\bibnamefont {Kosloff}},\ }\href@noop
  {} {\bibfield  {journal} {\bibinfo  {journal} {Entropy}\ }\textbf {\bibinfo
  {volume} {18}},\ \bibinfo {pages} {124} (\bibinfo {year} {2016})}\BibitemShut
  {NoStop}%
\bibitem [{\citenamefont {Brand{\~a}o}\ \emph {et~al.}(2015)\citenamefont
  {Brand{\~a}o}, \citenamefont {Horodecki}, \citenamefont {Ng}, \citenamefont
  {Oppenheim},\ and\ \citenamefont {Wehner}}]{BrandaoPnasRT2ndLaw}%
  \BibitemOpen
  \bibfield  {author} {\bibinfo {author} {\bibfnamefont {F.}~\bibnamefont
  {Brand{\~a}o}}, \bibinfo {author} {\bibfnamefont {M.}~\bibnamefont
  {Horodecki}}, \bibinfo {author} {\bibfnamefont {N.}~\bibnamefont {Ng}},
  \bibinfo {author} {\bibfnamefont {J.}~\bibnamefont {Oppenheim}}, \ and\
  \bibinfo {author} {\bibfnamefont {S.}~\bibnamefont {Wehner}},\ }\href@noop {}
  {\bibfield  {journal} {\bibinfo  {journal} {Proceedings of the National
  Academy of Sciences}\ }\textbf {\bibinfo {volume} {112}},\ \bibinfo {pages}
  {3275} (\bibinfo {year} {2015})}\BibitemShut {NoStop}%
\bibitem [{\citenamefont {Vinjanampathy}\ and\ \citenamefont
  {Anders}(2016)}]{SaiJanetReview}%
  \BibitemOpen
  \bibfield  {author} {\bibinfo {author} {\bibfnamefont {S.}~\bibnamefont
  {Vinjanampathy}}\ and\ \bibinfo {author} {\bibfnamefont {J.}~\bibnamefont
  {Anders}},\ }\href@noop {} {\bibfield  {journal} {\bibinfo  {journal}
  {Contemporary Physics}\ }\textbf {\bibinfo {volume} {57}},\ \bibinfo {pages}
  {545} (\bibinfo {year} {2016})}\BibitemShut {NoStop}%
\bibitem [{\citenamefont {Funo}\ and\ \citenamefont
  {Ueda}(2015)}]{FunoWorkFlucTradeOffRenyi}%
  \BibitemOpen
  \bibfield  {author} {\bibinfo {author} {\bibfnamefont {K.}~\bibnamefont
  {Funo}}\ and\ \bibinfo {author} {\bibfnamefont {M.}~\bibnamefont {Ueda}},\
  }\href@noop {} {\bibfield  {journal} {\bibinfo  {journal} {Physical review
  letters}\ }\textbf {\bibinfo {volume} {115}},\ \bibinfo {pages} {260601}
  (\bibinfo {year} {2015})}\BibitemShut {NoStop}%
\bibitem [{\citenamefont {Wilming}\ and\ \citenamefont
  {Gallego}(2017)}]{wilming2017ThirdLawRT}%
  \BibitemOpen
  \bibfield  {author} {\bibinfo {author} {\bibfnamefont {H.}~\bibnamefont
  {Wilming}}\ and\ \bibinfo {author} {\bibfnamefont {R.}~\bibnamefont
  {Gallego}},\ }\href@noop {} {\bibfield  {journal} {\bibinfo  {journal} {arXiv
  preprint arXiv:1701.07478}\ } (\bibinfo {year} {2017})}\BibitemShut {NoStop}%
\bibitem [{\citenamefont {Woods}\ \emph {et~al.}(2015)\citenamefont {Woods},
  \citenamefont {Ng},\ and\ \citenamefont {Wehner}}]{Woods2015EffEng}%
  \BibitemOpen
  \bibfield  {author} {\bibinfo {author} {\bibfnamefont {M.~P.}\ \bibnamefont
  {Woods}}, \bibinfo {author} {\bibfnamefont {N.}~\bibnamefont {Ng}}, \ and\
  \bibinfo {author} {\bibfnamefont {S.}~\bibnamefont {Wehner}},\ }\href@noop {}
  {\bibfield  {journal} {\bibinfo  {journal} {arXiv preprint arXiv:1506.02322}\
  } (\bibinfo {year} {2015})}\BibitemShut {NoStop}%
\end{thebibliography}%

\end{document}